\documentclass[a4paper,twocolumn,11pt,accepted=2022-04-26]{quantumarticle}

\pdfoutput=1
\usepackage[utf8]{inputenc}
\usepackage[english]{babel}
\usepackage[T1]{fontenc}
\usepackage{amsmath}
\usepackage{hyperref}
\usepackage{dsfont}
\usepackage{enumitem}
\usepackage{braket}
\usepackage{amsmath}
\usepackage{amssymb}
\usepackage{mathtools}
\usepackage{graphicx}
\usepackage{xcolor}
\usepackage{algorithm}
\usepackage{algpseudocode}
\usepackage{tikz}
\usepackage{lipsum}
\usepackage[font={footnotesize,sf}]{caption} 
\usepackage[numbers]{natbib}

\DeclarePairedDelimiter{\floor}{\lfloor}{\rfloor}
\newcounter{question}
\newcommand\Q[1]{%
   \leavevmode\par
  \stepcounter{question}
   \noindent
   \textbf{Q\thequestion:} #1}

\newcommand\A[2][]{%
    \leavevmode\par\noindent
   {\textbf{A\thequestion:} \textbf{#1}#2\par}}

\begin{document}

\title{Re-examining the quantum volume test: Ideal distributions, compiler optimizations, confidence intervals, and scalable resource estimations}

\author{Charles H. Baldwin}
\email{charles.baldwin@quantinuum.com}
\orcid{0000-0003-3938-7428}
\author{Karl Mayer}
\author{Natalie C. Brown}
\author{Ciar\'an Ryan-Anderson}
\author{David Hayes}
\affiliation{Quantinuum, 303 S. Technology Ct, Broomfield, Colorado 80021, USA}

\begin{abstract}
The quantum volume test is a full-system benchmark for quantum computers that is sensitive to qubit number, fidelity, connectivity, and other quantities believed to be important in building useful devices. The test was designed to produce a single-number measure of a quantum computer's general capability, but a complete understanding of its limitations and operational meaning is still missing. We explore the quantum volume test to better understand its design aspects, sensitivity to errors, passing criteria, and what passing implies about a quantum computer. We elucidate some transient behaviors the test exhibits for small qubit number including the ideal measurement output distributions and the efficacy of common compiler optimizations. We then present an efficient algorithm for estimating the expected heavy output probability under different error models and compiler optimization options, which predicts performance goals for future systems. Additionally, we explore the original confidence interval construction and show that it underachieves the desired coverage level for single shot experiments and overachieves for more typical number of shots. We propose a new confidence interval construction that reaches the specified coverage for typical number of shots and is more efficient in the number of circuits needed to pass the test. We demonstrate these savings with a $QV=2^{10}$ experimental dataset collected from Quantinuum System Model H1-1. Finally, we discuss what the quantum volume test implies about a quantum computer's practical or operational abilities especially in terms of quantum error correction.
\end{abstract}

\maketitle

\tableofcontents

\section{Introduction}
Quantum computers continue to advance towards higher performance devices that are nearing the regime of running advantageous algorithms. However, with several different device architectures and candidate algorithms, an open question remains: how do we quantify performance? The quantum volume (QV) metric was originally proposed as an answer to this question by weighing qubit number with fidelity~\cite{Bishop17, Moll18}, or simply stated as, ``don't count your qubits until you can entangle them''~\footnote{The original version of this phrase seems to have come from Robert Sutor in a talk given at Vanderbilt University entitled, \href{https://www.robertsutor.com/2019/02/24/talk-quantum-computing-dont-count-your-qubits-before-theyre-hatched/}{``Don't count your qubits until they hatch''}}. Later, QV was formalized in an explicit set of test circuits and passing criteria~\cite{Cross19}, which we refer to as the quantum volume test (QVT), and has recently been measured on several quantum computers~\cite{Cross19, Pino21, Sundaresan20, Jurcevic21, HQS128, IBM128, HQS512, HQS1024}. In this paper, we present a detailed study of the test for arbitrary qubit number $N$, referred to as $\textrm{QVT}_N$.

The QVT is an example of a broadening focus in quantum computer benchmarking from the component level to the system level. Component-level benchmarks, e.g., tomography~\cite{Chuang97} and randomized benchmarking~\cite{Knill08, Magesan12}, return rigorous estimates (with some assumptions) of the primitive components, e.g., fidelity of state preparation, gates, and measurement. While widely used for building quantum computers, component-level benchmarks often fail to describe the behavior of larger quantum circuits~\cite{Lubinski21, Proctor20}. This may be due to additional errors that are missed by component-level benchmarks~\cite{Sarovar20}. System-level benchmarks are designed to be sensitive to such errors, and therefore provide valuable feedback on the feasibility of running larger circuits on a quantum computer. Some system-level benchmarks have adopted component-level techniques to estimate the fidelity of full-system operations~\cite{Proctor20, Boixo18, Erhard19, Proctor19, Harper20}. Other system-level benchmarks --- like QVT --- abandon the expressed goal of measuring errors and instead look to demonstrate that the system passes performance criteria deemed to be ``hard''~\cite{Lubinski21,Wright19,Cornelissen21}. 

The value of different system-level benchmarks is beyond the scope of this paper, but whatever opinion one might have, it is self-evident that correctly interpreting any benchmark result requires an in-depth understanding of the test. This calls for a clear analysis at several levels: (1) the motivation and consequences of design decisions used to build the test, (2) how the protocol responds noise, and (3) how the performance metric relates to other useful tasks in quantum information processing. The original QVT proposal in Ref.~\cite{Cross19} analyzed most of these tasks to motivate the use of the test. In this work, we expand on all points by performing a series of analytic and numerical studies of $\textrm{QVT}$ to better understand experimental test results and inform future performance goals.

We briefly discuss a few results of our study here. First, QVT circuits' ideal behavior and the effectiveness of compiler optimizations are functions of $N$ (including whether $N$ is even or odd). Second, success in QVT is mostly scales to the total gate error magnitude and not the source of errors. Third, the confidence interval proposed in Ref.~\cite{Cross19} is more restrictive than necessary, and we define a new confidence interval method that allows fewer circuits to reach the desired confidence level. Finally, the required gate fidelity to pass $\textrm{QVT}_N$ for near term devices aligns reasonably well with other near-term goals such as early demonstrations of quantum error correction.

This paper is organized as follows: In Sec.~\ref{sec:overview} we review the basic steps in QVT. Next, in Sec.~\ref{sec:faq} we answer some frequently asked questions about QVT and refer to later sections for more detail. Then, in Sec.~\ref{sec:circuits} we analyze the ideal behavior of the QVT$_N$ circuits and different effects of previously proposed compiler optimizations. In Sec.~\ref{sec:numerics} we perform numerical simulations to estimate QVT$_N$ success probabilities under different error models and predict future error targets with a scalable method. In Sec.~\ref{sec:confidence} we study the confidence intervals for QVT$_N$ and propose a new method with tighter coverage. In Sec.~\ref{sec:comparisons} we compare QVT$_N$ results to other algorithms such as quantum error correction. Finally, in Sec.~\ref{sec:conclusions} we summarize our work and discuss open questions.

\section{Overview of the Quantum Volume Test} \label{sec:overview}
In Ref.~\cite{Cross19}, Cross \textit{et al.} outlined the QVT$_N$ procedure, which we summarize below. The task of QVT$_N$ is to experimentally run a type of random quantum circuit and generate output distributions exhibiting characteristics of a random unitary ensemble. This is quantified by a measure called the heavy output frequency (defined below). The procedure was inspired by Ref.~\cite{Aaronson16}, which proposed methods to demonstrate quantum computational advantage in sampling, where they asserted that there is no polynomial-time classical method that samples heavy outputs at least 2/3 of the time (under several assumptions). Therefore, observing heavy outputs more than 2/3 of the time from a quantum computer is an indication of a quantum speedup in sampling. 

\begin{figure} 
  \includegraphics[width=\columnwidth]{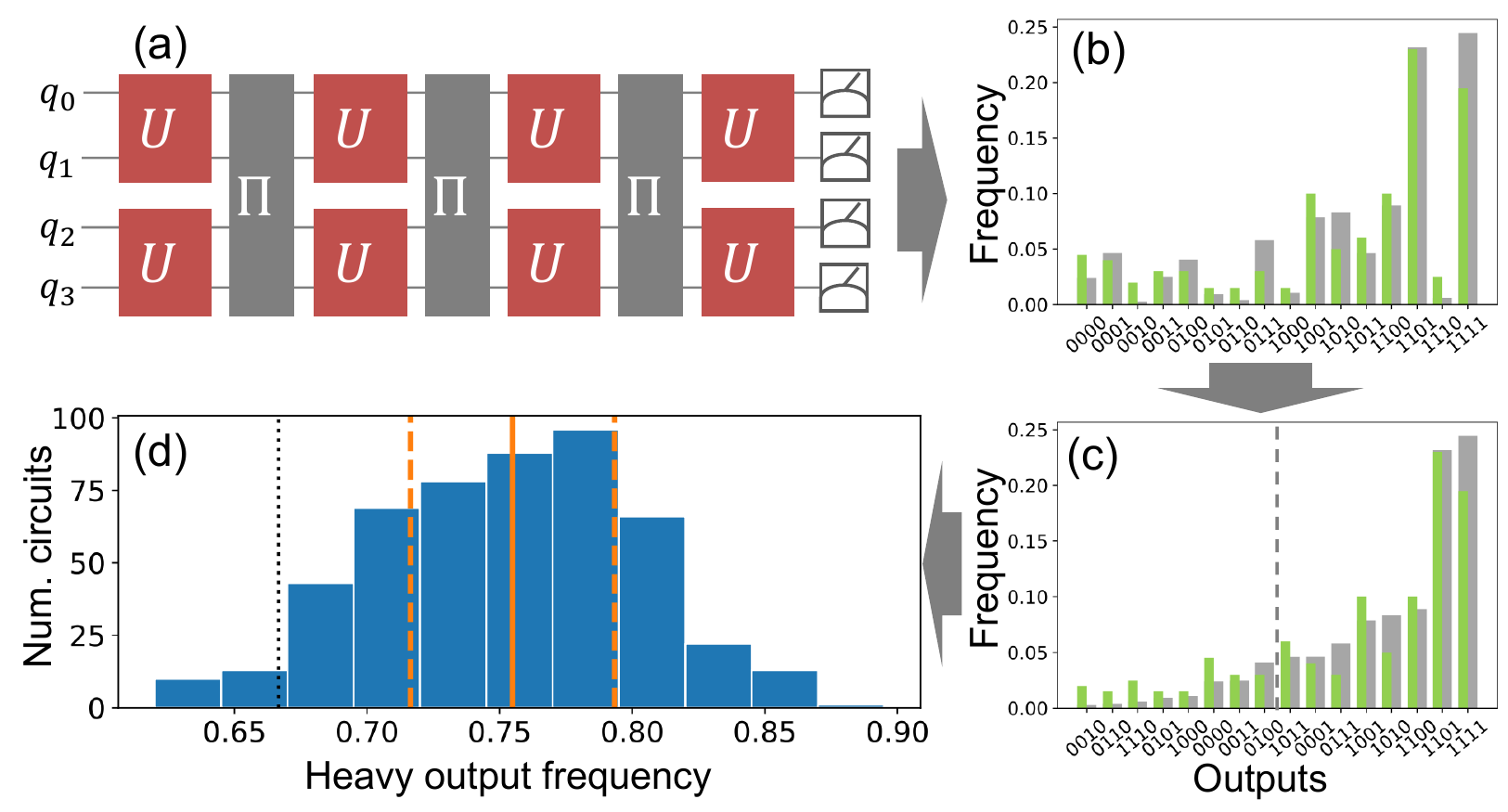}
  \caption{Steps in QVT$_N$. (a) A QVT$_4$ circuit consisting of alternating layers of random SU(4) gates (depicted as $U$'s in the circuit) acting on pairs of qubits followed by random permutations ($\Pi$) of qubits for different pairings in the next round. (b) The circuit is run several times on the quantum computer to estimate the resulting measurement distribution (illustrated by green histograms) and classically simulated to generate the ideal distribution (gray histogram). Here, the different measurement outputs are labeled by the bit strings on the x-axis, ordered in the standard binary system. (c) The ideal probabilities generated in the classical simulation are sorted in increasing order, so the least probable measurement output is on the left. The heavy outputs are labeled by bit strings whose output probabilities are greater than the median of the rearranged distribution. (d) The process is repeated for $n_c \geq 100$ circuits and the heavy output frequency distribution is plotted (blue histograms). When the average heavy output frequency (solid orange line) is $>2/3$ (dashed black line) with 97.73\% (or two-sigma) confidence (dashed orange line), the quantum computer has passed the test and said to have a $QV = 2^N$ ($16$ in the illustrated case).} 
\label{fig:qv_circuits}
\end{figure}

In general, $\textrm{QVT}_N$ is performed by running $n_c\geq 100$ different random quantum circuits on the quantum processor under investigation, and certifying their performance with classical simulation. As an example, the procedure for QVT$_4$ is outlined in Fig.~\ref{fig:qv_circuits}. The circuits are constructed by randomly pairing qubits and applying Haar-random SU$(4)$ gates to each pair as shown in Fig.~\ref{fig:qv_circuits}a, (for odd $N$ one qubit is left out in each of these rounds). The random pairing and gating is repeated $N$ times for $N$ qubits making the circuits ``square,'' since the depth (number of non-parallel gates) is on the order of the width (qubit number). Each circuit is simulated classically to determine the ideal distribution of measurement outputs in the standard computational basis (Fig.~\ref{fig:qv_circuits}b). The simulated distribution is then sorted according to the relative ideal probabilities of each output and the median output is found (Fig.~\ref{fig:qv_circuits}c). Heavy outputs are defined as measurement outputs with an ideal probability greater than the median. Each circuit is then run $n_s$ times on the device and the ratio of heavy outputs observed to the total shots in the experiment $n_s \times n_c$ is calculated and called the heavy output \textit{frequency} $\hat{h}$. The confidence interval lower bound of $\hat{h}$ is estimated as
\begin{equation} \label{eq:confidence_ibm}
C_{\textrm{lower}} = \hat{h} - 2 \sqrt{\frac{\hat{h} (1 - \hat{h})}{n_c}},
\end{equation}
which is derived assuming all circuits have the same number of shots. If $C_{\textrm{lower}} > 2/3$, QVT$_N$ is passed and the system has $QV = 2^N$.

Throughout this article, we study the heavy output probability averaged over the set of $all$ possible QVT$_N$ circuits. Without errors, we call this quantity the ideal success $h_{\textrm{ideal}}$. With errors, we call this quantity the actual success $h$. The actual success does not include finite sampling of the data to differentiate those effects from errors. In general errors cause $h \leq h_{\textrm{ideal}}$. 

\section{FAQ's about the QVT} \label{sec:faq}
Since this paper covers a wide range of topics, we first attempt to answer some frequently asked questions about QV and refer the interested reader to more details in the corresponding sections below.

\Q{What is the average heavy output probability without errors?}
\A{In the original QVT proposal~\cite{Cross19}, the asymptotic ideal heavy output probability is given as $\approx 84.7\%$. This means QVT$_N$ success is not equivalent to fidelity, which equals one without errors. In Sec.~\ref{sec:comparisons} we propose a scaling method to better interpret QVT$_N$ measurements that relates more closely to circuit fidelity. It was also demonstrated in Ref.~\cite{Cross19} that circuits with small $N$ have slight deviation of heavy output probabilities from the asymptotic value. In Sec.~\ref{sec:ideal} we shed additional light on this deviation and we find that for $N<10$ ideal success varies with qubit number by about 1-2\%. This may seem like a small variation but in practice could mean the difference between passing and not passing. We also find a difference in scaling of ideal heavy output probability for odd $N$ vs even $N$. This means that the success between dimensions is difficult to compare. For example, heavy output probability of 70\% for QVT$_2$ may be require lower errors than 70\% for QVT$_3$ because the ideal heavy output probability for QVT$_3$ is much higher.}

\Q{The QVT allows arbitrary compiler optimizations (within reason~\cite{Cross19}), but what effect do they have on the test?}
\A{Classical compilation plays an important role in NISQ algorithms, especially on machines with limited connectivity and the QVT rewards quantum compilers' ability to optimize circuit compositions. Ref.~\cite{Cross19} proposed two optimizations for QVT that reduce the total number of two-qubit gates to improve the chances of success. We find these optimizations help significantly for $N \leq 10$ qubits, for example reducing the number of two-qubit gates by about half for $N=4$, but provide diminishing advantages as $N$ increases, for example only a 20\% reduction for the same methods with $N=15$. We explore the exact scaling to better determine advantages for any qubit number in Sec.~\ref{sec:optimizations}.}

\Q{How does $\textrm{QVT}_N$ success scale with two-qubit gate fidelity?}
\A{Most systems are limited by two-qubit gate errors, making them a primary focus in running any benchmark or algorithm. We find that the success of QVT$_N$ experiments roughly scales with the fidelity of two-qubit gates $f_{TQ}$ raised to the expected number of two-qubit gates $n_{TQ}$, $h \propto f_{TQ}^{n_{TQ}}$. The total number of two-qubit gates is at most $3 \floor{N/2} N$ (where ``floor $m$'' $\floor{m}$ rounds $m$ down to the nearest integer) but can be significantly reduced for $N\leq 10$ qubits with compiler optimizations (see Sec.~\ref{sec:circuits}). Also this scaling does not take into account other error sources like single-qubit gate, measurement errors or crosstalk errors, which also impact QVT$_N$ success. This scaling is only a rough estimate and a more detailed analysis is presented in Sec.~\ref{sec:numerics}.}

\Q{Is $\textrm{QVT}_N$ only sensitive to two-qubit gate error?}
\A{Two-qubit gate errors are the main concern for most systems, but QVT$_N$ also requires single-qubit gates and of course state preparation and measurement as well as being sensitive to other errors like crosstalk and idling errors. For single-qubit gate fidelity $f_{\textrm{SQ}}$, we find that a similar expression holds as in the previous question $h \propto f_{SQ}^{2 n_{TQ} + N}$ since there are roughly two single-qubit gates for every two-qubit gate plus $N$ additional gates at the beginning of each circuit. For state preparation and measurement we observe a softer exponential scaling with fidelity $f_{P/M}$ since there are only $N$ state preparations and measurements $h \propto f_{P/M}^{N}$. Similar to Q3, these scalings are rough estimates. A more detailed method that can account for other errors and combinations of multiple errors is presented in Sec.~\ref{sec:numerics} and compared to numerical simulation. We also attempt to simulate effects like crosstalk but of course these are system specific, and therefore it is important to run QVT$_N$ on actual hardware to demonstrate low levels of errors.}

\Q{Does QVT$_N$ have different behavior with different types of errors?}
\A{We find that QVT$_N$ behaves similarly with different types errors of similar magnitudes as measured by infidelity. This is best exemplified by comparing two-qubit coherent errors to depolarizing errors. We find both of these error models produce similar QVT$_N$ success when they have the same gate fidelities in Sec.~\ref{sec:numerics}. We observe similar trends for other error models as well. It is impossible to simulate all possible errors but we expect QVT$_N$ success to be mostly a simple function of fidelity rather than depending on the type of error.}

\Q{QVT requires classical simulation in the analysis, doesn't this put a limit on the usefulness of the test?}
\A{For $N>30$ the QVT will be difficult to implement since the classical computation will be expensive. As estimated in Sec.~\ref{sec:numerics}, passing $\textrm{QVT}_{30}$ likely requires a two-qubit gate fidelity of $\approx 99.95\%$ along with low single-qubit gate errors and minimal crosstalk and memory errors. As studied in Sec.~\ref{sec:comparisons}, reaching these performance levels with 30 qubits is a worthy medium-term goal for developing quantum computing platforms with a variety of applications. Moreover, failure to run QVT due to the inability to classically simulate the system dynamics implies the system has achieved quantum sampling advantage, which is a good problem to have.}

\Q{How reasonable is the passing criteria for QVT$_N$?}
\A{The passing criteria for QVT$_N$ is to observe an average heavy output frequency above 2/3 with two-sigma confidence. The passing criteria of 2/3 chosen from Ref.~\cite{Aaronson16} and was used for proofs of quantum advantage. For reference, without errors the highest possible heavy output frequency we expect is $\approx 84.7\%$ for asymptotically large $N$ and the lowest is 1/2 for completely depolarizing circuits of any $N$. We find in Sec.~\ref{sec:confidence} that the confidence intervals constructed in the original proposal~\cite{Cross19} are much wider than necessary to achieve the specified two-sigma coverage. We propose a new method for constructing confidence intervals that provides tighter bounds with the specified coverage probability and we validate the method with numerical tests. In Sec.~\ref{sec:qec} we run simulations to compare the estimated gate fidelity needed to pass QVT$_N$ to the estimated gate fidelity needed to cross the pseudo-thresholds for different small-distance quantum error correction codes. We find that gate fidelity necessary to pass QVT$_N$ for larger $N$ corresponds to circuit fidelity that is much larger than what is necessary quantum advantage demonstrations~\cite{Arute19}. However, the gate fidelity needed for QVT$_N$ is reasonably in-line with achieving break-even QEC, and thereby enabling large-scale computations.}

\section{Circuits} \label{sec:circuits}
In this section we explore QVT circuit construction and optimization to better understand and predict the heavy output frequencies in experiments. QVT specifies a circuit construction method (outlined in Sec.~\ref{sec:overview}) in an attempt to generate output distributions that are typical of random quantum circuits. After generating the circuits, QVT allows any circuit compilations that leave the net unitary ``close'' to the original ideal unitary (further specified below). Two methods that satisfy this condition were proposed in Ref.~\cite{Cross19}. We propose an additional method and study how these methods scale for arbitrary qubit number and fidelity.

\subsection{Ideal distribution} \label{sec:ideal}
In previous work it was shown that circuits generated with random two-qubit gates on pairs of qubits (like those in QVT) form approximate unitary $t$-designs if sufficiently deep~\cite{Oliveira07, Harrow09, Brandao16}. Unitary $t$-designs approximate the first $t$ moments of the Haar measure, which is the invariant measure across unitaries of fixed dimension~\cite{Dankert09}. Here, we study the output states of Haar random unitaries and compare them to the output states from QVT circuits.

\begin{figure} 
  \includegraphics[width=\columnwidth]{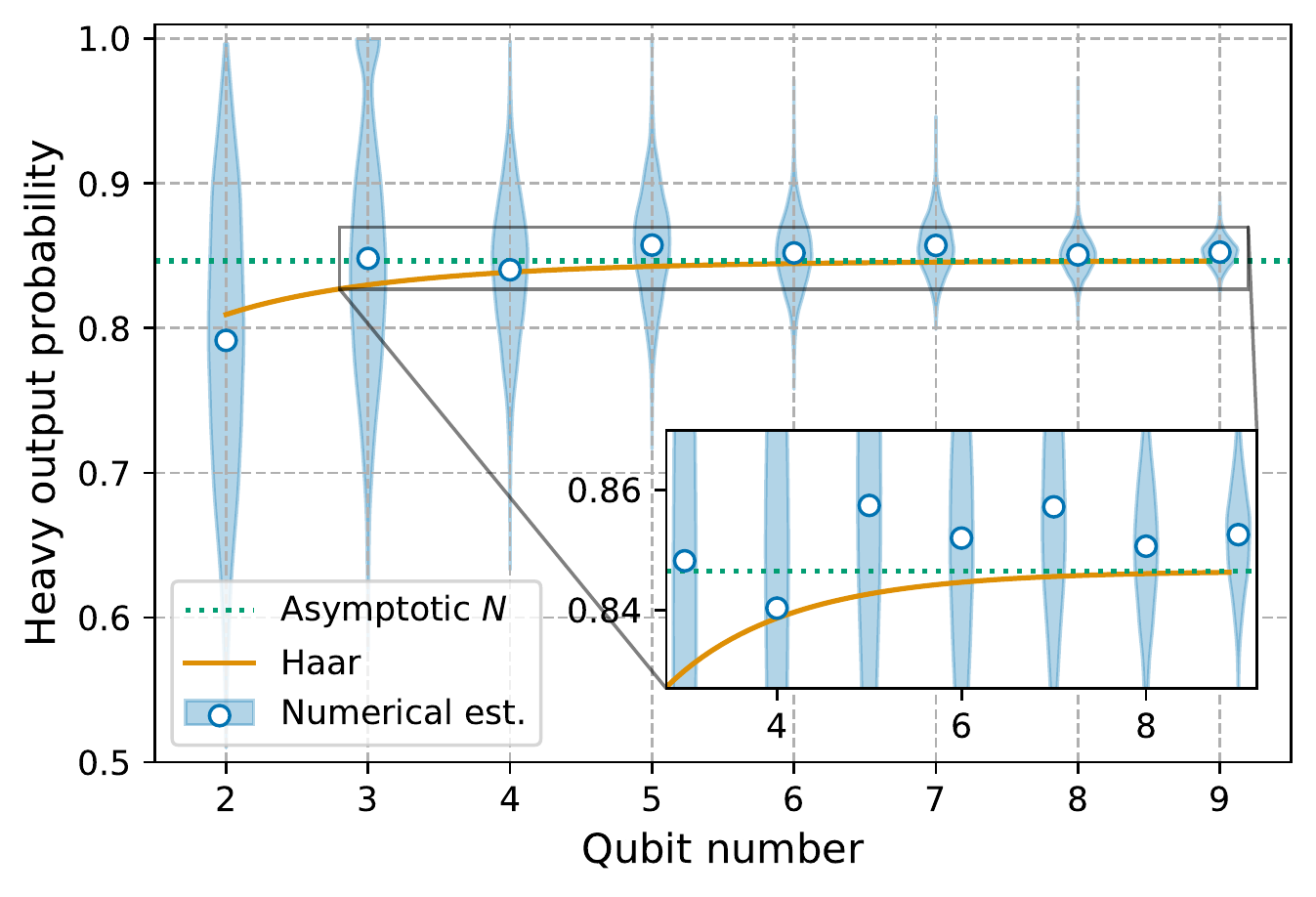}
  \caption{The ideal success as a function of $N$ estimated from sampling 5,000 QVT circuits for each $N$. Blue regions give the distribution of heavy output probabilities defined by individual circuit medians over the sample circuits and blue circles show the average. The orange solid line shows the expected heavy output probability over all Haar random SU$(2^N)$ unitaries from Eq.~\eqref{eq:Haar_exp}. The green dashed line is the asymptotic limit of the PT distribution, $ (1 + \textrm{log}2)/2\approx 84.7\%$. }
\label{fig:ideal distribution}
\end{figure}

First, we derive the expected heavy output probability for Haar random unitaries. A Haar random state is generated from applying a Haar random unitary to any initial state. Haar random states are a superposition of computational basis states $\ket{\psi} = \sum_{j=1}^{2^N} c_j \ket{j}$ for amplitudes $c_j$, with real and imaginary parts uniformly distributed between [-1, 1] subject to the normalization condition. The probability of measuring each computational basis output $x_k$ is $p(x_k) =| \braket{x_k|\psi} |^2= |c_k|^2$. The probability distribution of $p(x_k)$ is found by integrating over the Haar measure $\mathcal{P}_{H}(p) = (2^N - 1) (1 - p)^{2^N - 2}$~\cite{Boixo18, Mullane20}. This is a probability distribution over output probabilities averaged over all Haar random states of dimension $2^N$. The expected heavy output probability is derived by finding the median probability of the distribution and then integrating over all probabilities above the median
\begin{equation} \label{eq:Haar_exp}
	h_{\textrm{ideal}}(N) = 2^{2^N/(1 - 2^N)}  (1 + 2^N (2^{1/(2^N - 1)} - 1)).
\end{equation}

For large $N$, $\mathcal{P}_{H}(p)$ approaches the Porter-Thomas (PT) distribution $\mathcal{P}_{PT}(p) = 2^N e^{-p 2^N}$. The large $N$ approximations leads to the asymptotic ideal success probability of QVT circuits of $h_{\textrm{ideal}} \approx (\log 2 + 1)/2\approx 84.7\%$~\cite{Cross19, Aaronson16}.

\begin{figure*}[]
	\begin{center}
		\includegraphics[width=\textwidth]{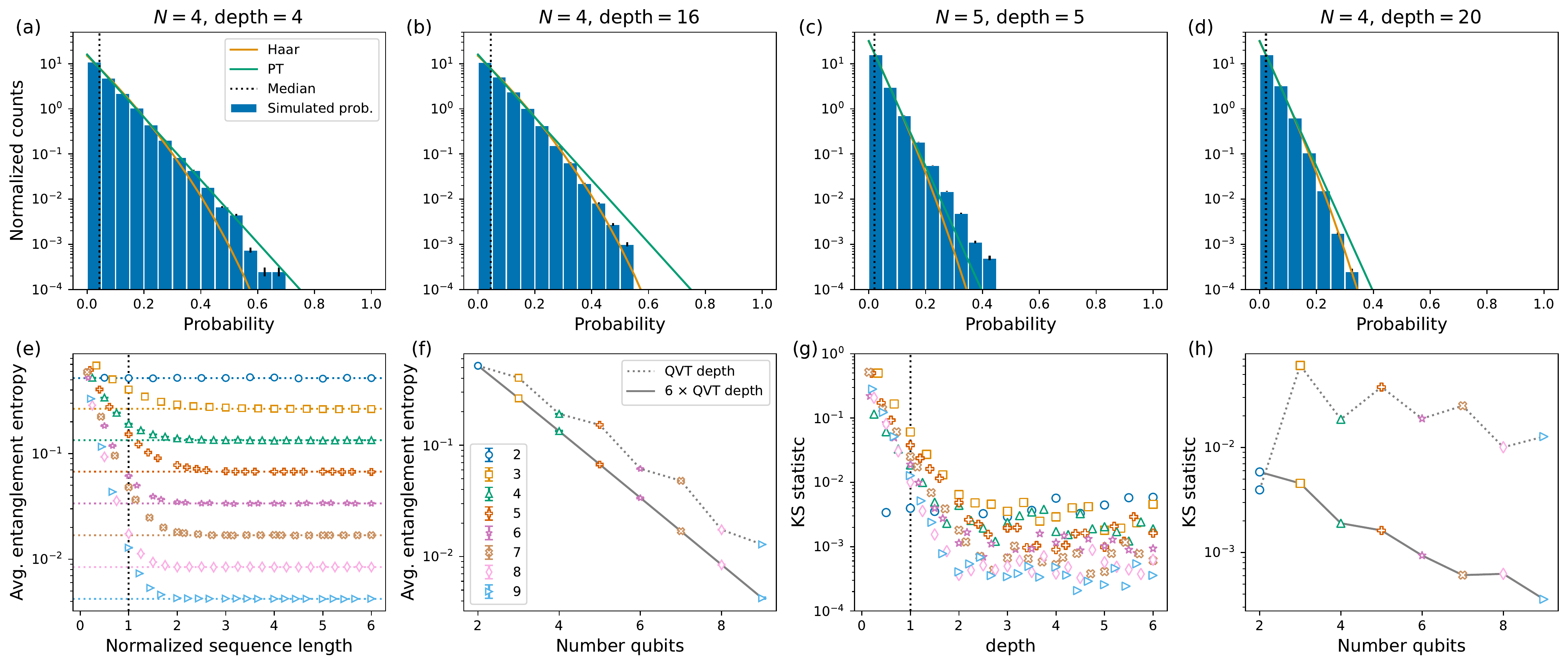}
		\caption{(Top row) Example probability distributions from 5,000 random QVT$_N$ circuits of different $N$ and depth. Blue histograms show numerical results from sample circuits. Black lines show estimated errorbars from multinomial distribution. Green line is the asymptotic estimate and orange line is the expected distribution from Haar random SU($2^N$). From left to right: (a) $N=4$, depth$=4$ (standard QVT$_4$), (b) $N=4$, depth$=16$,  (c) $N=5$, depth$=5$ (standard QVT$_5$),  (d) $N=5$, depth$=20$. (Bottom row) Comparisons of output distribution from the sample of 5,000 QVT$_N$ circuits for various $N$ as a function of depth (normalized by $N$). (e) Average entanglement entropy across all single-qubit partitions (between 2-dimensional and $(2^{N-1}$-dimensional Hilbert spaces). Colored dashed lines are the expected limit derived in Ref.~\cite{Page93}. (f) Slice over $N$ depth circuits (dashed line) corresponding to standard QVT$_N$ test and $6 N$ depth circuits (solid line). (g) KS test statistic between binned probabilities from 5,000 circuit sample and the expected Haar distribution. (h) Slice over $N$ depth circuits (dashed line) corresponding to standard QVT$_N$ test and $6 N$ depth circuits (solid line).}
		\label{fig:ideal scaling}
	\end{center}
\end{figure*}

In Fig~\ref{fig:ideal distribution} we plot a comparison between different estimates of the expected heavy output probability and the heavy output probability from a sample of 5,000 QVT circuits. The simulated data is plotted in blue regions to show the distribution of heavy output probabilities based on circuit instance with mean (estimated ideal success) plotted as blue dots. One notable feature is that the estimated ideal success depends on $N$ and oscillates between higher values for odd $N$ and lower values for even $N$ while converging to the asymptotic value. For $N < 5$ there is also a notable difference between the heavy output probability predicted by the Haar distribution $\mathcal{P}_{H}(p)$ (orange solid line) and the asymptotic estimate (dashed green line). There is also a discrepancy between the heavy output probability from the Haar distribution (orange solid line) and estimated ideal success from QVT$_N$ (blue circles). We partially attribute the oscillations and discrepancies to two reasons: First, the ideal success is calculated by estimating the median probability \textit{per circuit} whereas the derived result is estimated by the median over \textit{all outputs}, and second, the QVT$_N$ circuits are not representative of Haar random SU($2^N$) unitaries (and in fact the difference depends on odd versus even $N$). We further investigate the second point below.

To compare the output states from QVT circuits to Haar random states we conducted a numerical study of 5,000 QVT$_N$ circuits for $N=\{2,\dots,9\}$. We then extended the circuits with the same construction method out to $6N$ rounds of permutations and SU(4) gate pairings, which is $6 \times$ QVT$_N$ circuit depth. At various depths we extracted the quantum state for each circuit in order to study the how the output distributions converge. For each $N$ and circuit depth there are $5,000 \times 2^N$ probabilities and we see empirically that the corresponding distribution for small qubit number and depth do not match the PT or Haar distributions in the tails, e.g. Fig.~\ref{fig:ideal scaling}(a-d). To verify this observation, we conducted two tests. First, we calculated the average entanglement entropy over each qubit partition for each circuit. We traced out $N-1$ qubits in each circuit and calculated the entropy of the remaining subsystem and averaged over all qubits and circuits. Second, we applied the Kolmogorov-Smirnov (KS) test between the predicted Haar distribution and the simulated probabilities. The KS test measures the distance between a sampled distribution and a model probability distribution~\cite{NISTHandbook}. As shown in Fig~\ref{fig:ideal scaling}(e-h), both tests show that QVT$_N$ circuits do not produce the expected values for Haar random states but do converge with longer sequences, as expected based on Refs.~\cite{Harrow09, Brandao16}. For the KS test, the test statistic asympototes with circuit depth due to finite sampling effects, which are reduced for higher $N$ since there are more probabilities to compare. The finite sampling asymptote is not reached with the standard QVT$_N$ circuit depth, but is fairly close for 2$\times$ QVT$_N$ circuit depth. 

One other notable feature of the study is that the estimates of each test have higher entanglement entropy and KS test statistic for odd $N$ than for even $N$, indicating that odd $N$ circuits are further from SU($2^N$). One reason this occurs is that for odd $N$ some circuits have 100\% ideal heavy output probability, which is seen in Fig~\ref{fig:ideal distribution} in the violin plots. This occurs with 572/5,000 random circuits for $N=3$, 9/5,000 for $N=5$, and 0/5,000 for larger $N$. The reason is that for odd $N$ circuits one qubit is always left out per round, which means that in some circuits one qubit will be left out for all rounds. Then, the left out qubit totally determines the heavy outputs, which are outputs with the left out qubit in the $\ket{0}$ state. We can calculate the probability of sampling such a circuit based on the probability a qubit is left out in any given round. Since the pairings are random, after the initial round the probability that the same qubit is left out in the next round is $1/N$. Repeat this for all $N-1$ subsequent rounds that require repairing and the probability that the same qubit is left out every time is $1/N^{N-1}$.  This closely matches our numerical estimates: $N=3$ we expect 555.55 circuits, $N=5$ we expect 8 circuits, $N=7$ we expect $0.042$ circuits, and for $N = 9$ the expected circuits is $\leq 1.1 \times 10^{-4}$. This effect also diminishes quickly as $N$ increases, which matches our numerical comparisons in Fig.~\ref{fig:ideal scaling}.

\subsection{Compiler optimizations} \label{sec:optimizations}
In a QVT, any compilation method may be applied to the circuits such that the resulting unitary is close to the original unitary~\cite{Cross19}. One should not use the result of classical simulation in the compilation, e.g., finding the heavy outputs then designing the circuit that produces a single heavy output. Ref.~\cite{Cross19} proposed methods to compile the circuits to reduce the total number of two-qubit gates, and therefore improve the success. These compiler optimizations were not expected to scale favorably with qubit number and here we elucidate the exact scaling of two such optimizations: block combinations and block approximations. We also introduce a new optimization based on arbitrary angle gates.

\subsubsection{Block combinations} \label{sec:gate_combine}
\begin{figure}
	\centering
	\includegraphics[width=\columnwidth]{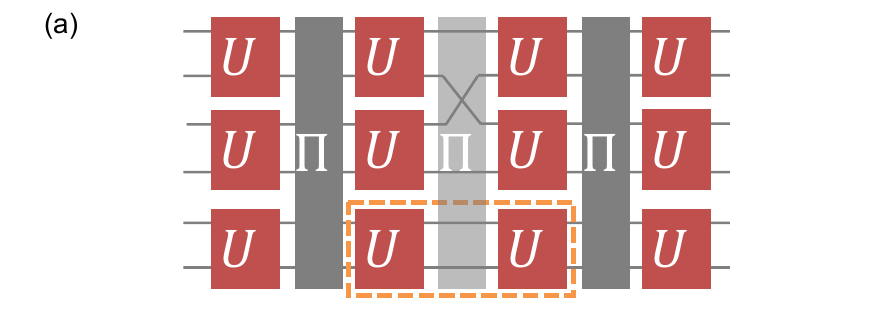}
	\includegraphics[width=\columnwidth]{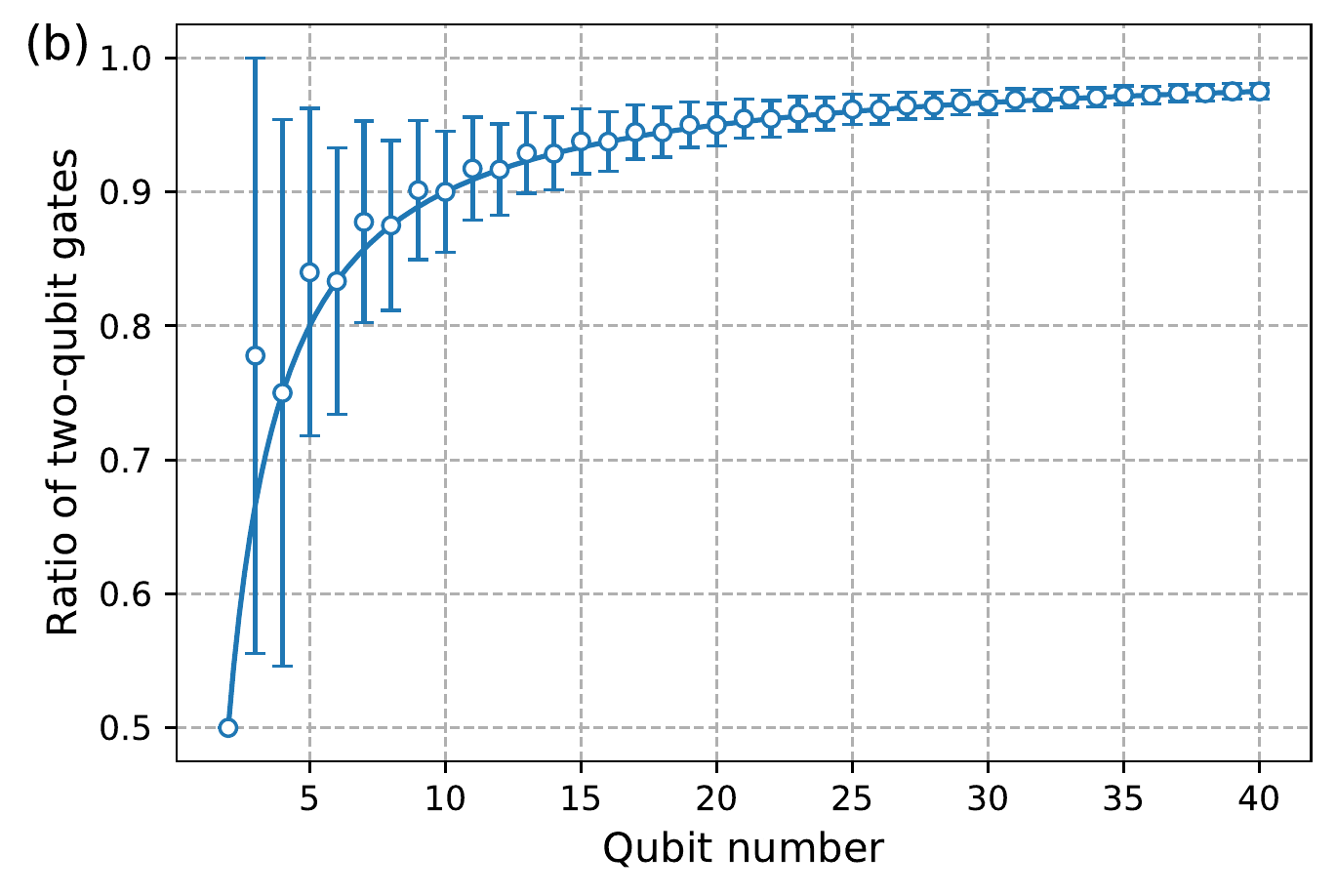}
	\caption{Overview of block combination and effectiveness. (a) Example block combine optimization where the second permutation $\Pi=(0,1,2,3,4,5) \rightarrow (0,1,3,2,4,5)$ leaves qubits 4 and 5 paired in the same way for the second and third rounds of SU(4) gates. Since the second and third SU$(4)$ gates acting on qubits 4 and 5 are random, it is equivalent to combine them into a single random SU$(4)$ gate, which should have a lower noise level than executing two gates, thereby increasing the success. (b) The average relative savings in the number of two-qubit gates upon taking advantage of block combinations. The blue line is equal to $(N-1)/N$, while the data points are found from Eq.~\eqref{eq:combined_gate}. The error bars shown are derived in a similar manner by solving for the standard deviation.}
	\label{fig:gate_combines}
\end{figure}

The block combination optimization takes $k \geq 2$ sequential blocks of SU$(4)$ gates scheduled to operate on the same two qubits and combines them into a single SU$(4)$ gate as shown in Fig.~\ref{fig:gate_combines}a. This reduces the number of two-qubit gates in this section of the circuit from $3k$ to 3. Here we calculate the average number of two-qubit gates saved as a combinatorial problem.

Each round of a QVT circuit requires the qubits to be divided into pairs that each receive random SU(4) gates. An arrangement represents this pairing for a given round and is defined by a set of tuples representing the paired qubits $\mathbf{p} = \{(0,1), (2,3),...\}$. We assume the first arrangement pairs the nearest neighbor qubits without loss of generality. Therefore, a QVT circuits contains a total of $N-1$ arrangements. 
 
The first step is to determine the total number of possible arrangements, denoted $f(N)$. For now, assume $N$ is even. Given an initial qubit, pick a second qubit to pair it with; there are $N-1$ choices. Iterate to the next qubit and pick its pair; there are $N-3$ remaining choices. The procedure continues until no qubits are remaining. The total number of possible arrangements is then $f(N) = (N-1)(N-3) \cdots 3 \cdot 1 = (N-1)!!$, where ``$!!$'' is a ``double factorial.'' By a similar argument for odd $N$, $f(N) = N!!$, and in general
\begin{align}
	f(N) &= \left.
		\begin{cases}
			N!! & \text{for } N = \text{ odd} \\
			(N-1)!! & \text{for } N = \text{ even} \\
		\end{cases}
	\right\} \nonumber \\
	 &=\frac{N!}{2^{\floor{N/2}}\floor{N/2}!}.
\end{align}
This subproblem is equivalent to finding the number of perfect matchings of a fully connected graph~\cite{Callan09}.

The next step is to find the number of times two consecutive rounds do not contain any repeated pairs, which we denote as $g(N)$. Let $S$ be the set of all possible arrangements and let $\mathbf{p}$ be the arrangement of the first round. Then let $S_{\mathbf{p}_j}$ represent the set of arrangements that contain the pairing $\mathbf{p}_j$, which is the $j$th pairing from the first round. Then the set of arrangements that do not repeat the pair $\mathbf{p}_j$ is the complement of $S_{\mathbf{p}_j}$ in the set $S$, which is denoted $\overline{S}_{\mathbf{p}_j}$. The set of arrangements with no repeats from the previous arrangement is then the intersection over all pairs $\mathbf{p}_j$ of the sets that do not contain that pair,
\begin{align}
	g(N) &= \left| \bigcap_{j=1}^{\floor{N/2}} \overline{S}_{\mathbf{p}_j} \right| \nonumber \\
	&= \left| \overline{\bigcup_{j=1}^{\floor{N/2}} S_{\mathbf{p}_j}} \right| \nonumber \\
	&= f(N) - \left|\bigcup \limits_{j=1}^{\floor{N/2}} S_{\mathbf{\mathbf{p}_j}}\right| \nonumber \\
	&= \sum_{k=0}^{\floor{N/2}} (-1)^{k} {\floor{N/2} \choose k} f(N - 2k),
\end{align}
where the second line uses De Morgan's law, and the last line uses the inclusion-exclusion principle~\cite{Roberts09} and noting that $|S_{\mathbf{p}_j}| = f(N-2)$, $|S_{\mathbf{p}_m} \cap S_{\mathbf{p}_{k\neq m}} | = f(N-4)$, etc.

Next, define the number of times two consecutive rounds contain exactly $M$ repeated pairs as $h(N,M)$. The $M$ repeated pairs are chosen in any combination from $\floor{N/2}$ pairs in the first round. The remaining $N-2M$ qubits must then contain no repeated pairs from the first round. Therefore,
\begin{equation}
h(N, M) = {\floor{N/2} \choose M} g(N - 2 M),
\end{equation}
which reduces to $g(N)$ for $M=0$ as expected.

The expected number of gates after opportunistic combining is $n_{\textrm{TQ}}(N)$ and is found by iterating through each round of the QVT circuit and calculating the fraction of circuits that require new pairs from the previous round. For the first round there are $3 \floor{N/2}$ gates since all pairs are new. In the next rounds, we iterate through the possible number of repeated pairs from the previous round $k$ from $k=0$ (no repeated pairs) to $k=\floor{N/2}$ (all repeated pairs). The fraction of total possible arrangements with exactly $k$ repeated pairs is $h(N,k)/f(N)$. For $k$ repeated pairs there are then $3(\floor{N/2} - k)$ new gates. This gives the expected total number of two qubit gates,
\begin{align} \label{eq:combined_gate}
	n_{\textrm{TQ}}(N) &= 3 \floor{N/2} \nonumber \\
	&+ \frac{3 (N -1)}{f(N)} \sum_{k=0}^{\floor{N/2}} h(N, k) (\floor{N/2} -k).
\end{align}

The expected fraction of gates saved with the block combinations $n_{\textrm{TQ}}(N)/(3 \floor{N/2} N)$ is plotted in Fig.~\ref{fig:gate_combines} along with standard deviations derived in a similar manner. For even qubit numbers we empirically see $n_{\textrm{TQ}}(N)/(3 \floor{N/2} N) = (N-1)/N$. Interestingly, the reduction is relatively less effective for odd $N$ than for $N+1$. This is because there are relatively fewer total pairs for odd $N$ compared to $N + 1$, and therefore less options to combine. In general, we find that the combine compilation roughly saves one round of the QVT$_N$ circuit.

\subsubsection{Block approximations} \label{sec:gate_approx}
\begin{figure} 
	\centering
	\includegraphics[width=\columnwidth]{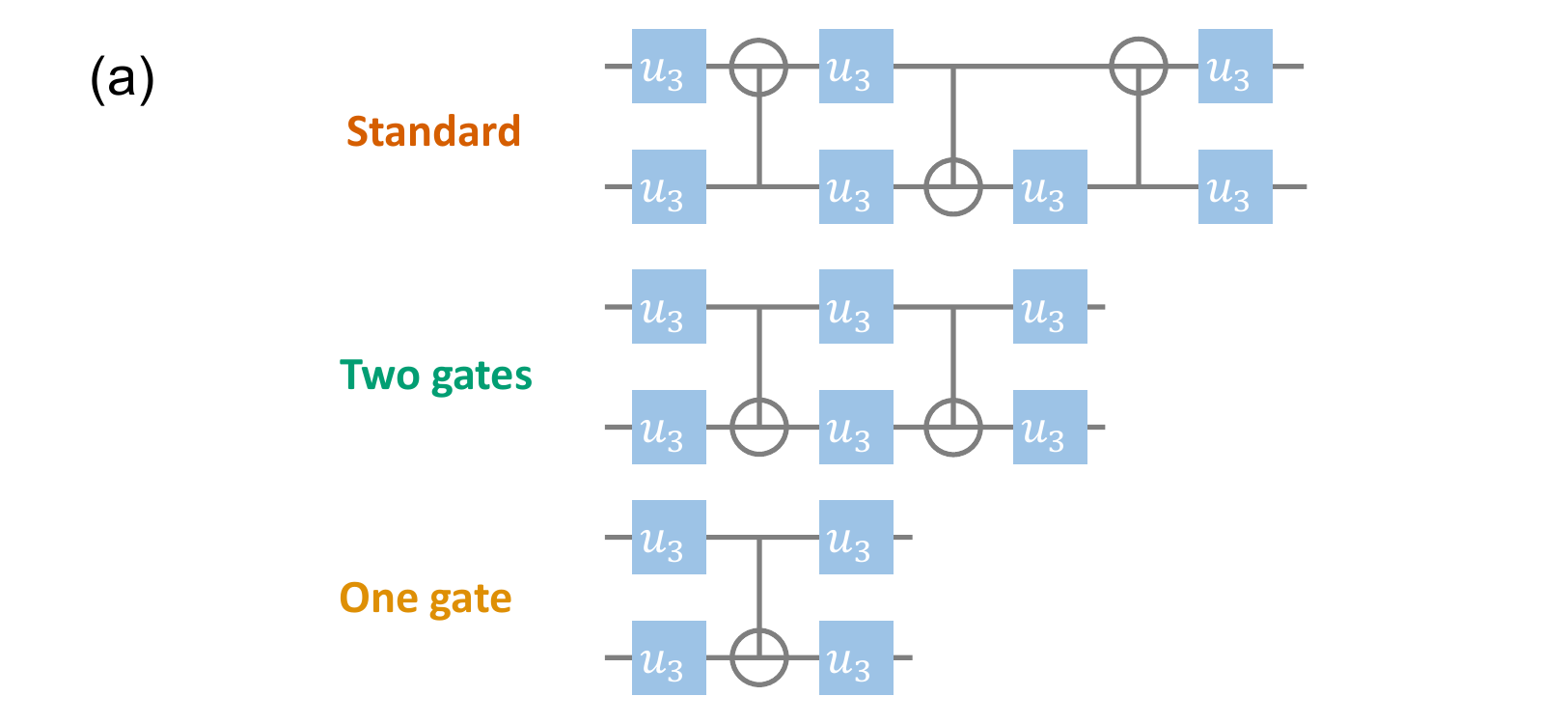}
	\includegraphics[width=\columnwidth]{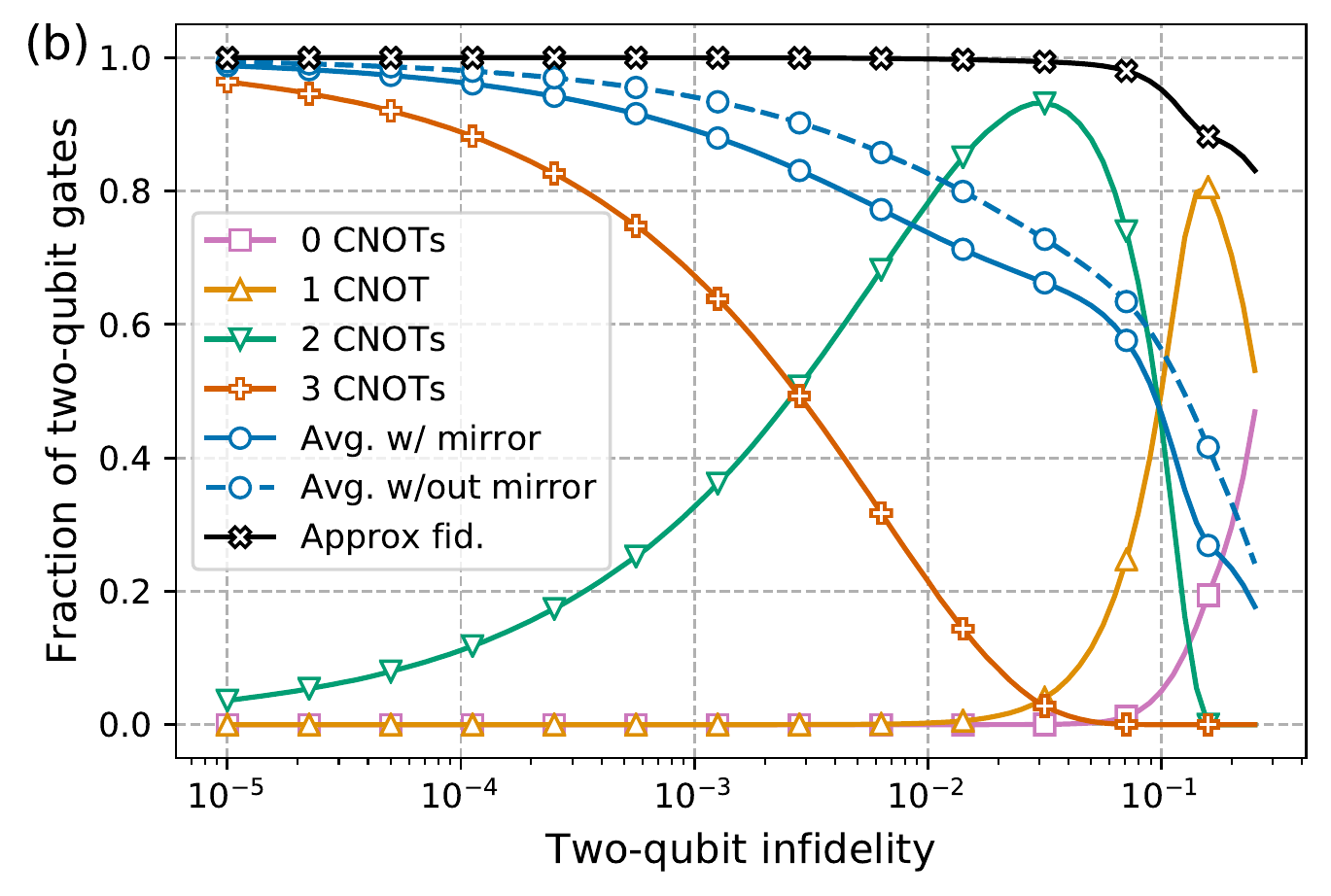}
	\caption{Numerical study of block approximation. (a) The gate decompositions used for generating exact and approximate random SU$(4)$ gates. Zero two-qubit gates (not shown) is two parallel single-qubit gates. (b) The fraction of random SU(4) gates that satisfy the given fidelity condition for replacement: (purple squares) zero two-qubit gates, (orange triangles) one two-qubit gate, (green up-side-down triangles) two two-qubit gates, and (red pluses) three two-qubit gates or no reduction. The average fraction of two-qubit gates are plotted in blue circles with mirroring option (solid line) and without mirroring option (dashed line). The fidelity of the approximation is plotted with black crosses.}
	\label{fig:gate_approx}
\end{figure}

The other compiling procedure proposed in Ref.~\cite{Cross19} is to replace the standard SU(4) block decomposition in Fig.~\ref{fig:gate_approx}a with an approximate version that contains fewer CNOT gates (or other perfect two-qubit entangler) if the approximate version has a higher estimated fidelity with errors. They also proposed a ``mirror'' option to additionally test SWAP$\times U$ to see if a corresponding approximation meets the fidelity conditions for replacement. Mirroring likely adds no overhead to the circuit, since all qubits are randomly reordered after the SU(4) block so on average no extra qubit routing is required. However, mirroring does change the SU(4) block potentially making it easier to implement with an approximation. More details are given in Ref.~\cite{Cross19}. If the condition is met for this mirror case then the new gate includes a SWAP and the qubit ordering is updated in future rounds to compensate. Ref.~\cite{Cross19} investigated both these options by deriving the fraction of SU(4) blocks that meet the fidelity criteria. Here, we extend this investigation to smaller gate error regimes of $10^{-1} -10^{-5}$ and numerically study performance with block combinations.

We performed a numerical search over 100,000~\footnote{\texttt{Qiskit} 0.28.0 only generates 1,000 random SU(4) blocks and samples from that set to generate every QVT circuit. In later simulations we do not restrict QVT circuits to using this smaller set. While QVT performance and current optimizations may not be impacted by using this smaller sample it is possible future optimization could use excessive classical computation on such a reduced set.} SU(4) blocks to determine what fraction meet the fidelity criteria with and without mirroring. The results are plotted in Fig.~\ref{fig:gate_approx} where the dashed blue line with circles shows the fraction of two-qubit gates returned from the approximation without mirroring and the solid blue line with circles shows the fraction of two-qubit gates with mirroring. The black line with crosses shows the fidelity of the resulting approximate gates without errors. The other colors show the fraction of different approximate versions of the SU(4) gates that meet the fidelity requirements. The orange curve with triangles shows a significant number of SU$(4)$ gates can be approximated with a single CNOT gate if an infidelity of $10^{-2}$ is acceptable, but for lower error rates there are very few. Likewise, the green curve with up-side-down triangles shows that even out to very low error rates of $10^{-4}$ a significant portion of random SU$(4)$ gates can be constructed using two CNOT operations. For large enough $N$, the $\textrm{QVT}_N$ will require an error rate $<10^{-4}$ at which point nearly all SU$(4)$ gates will require three CNOT's, but this is also beyond the regime where the current incarnation of the QVT is feasible due to the classical simulation requirement.

\begin{figure} 
	\centering
	\includegraphics[width=\columnwidth]{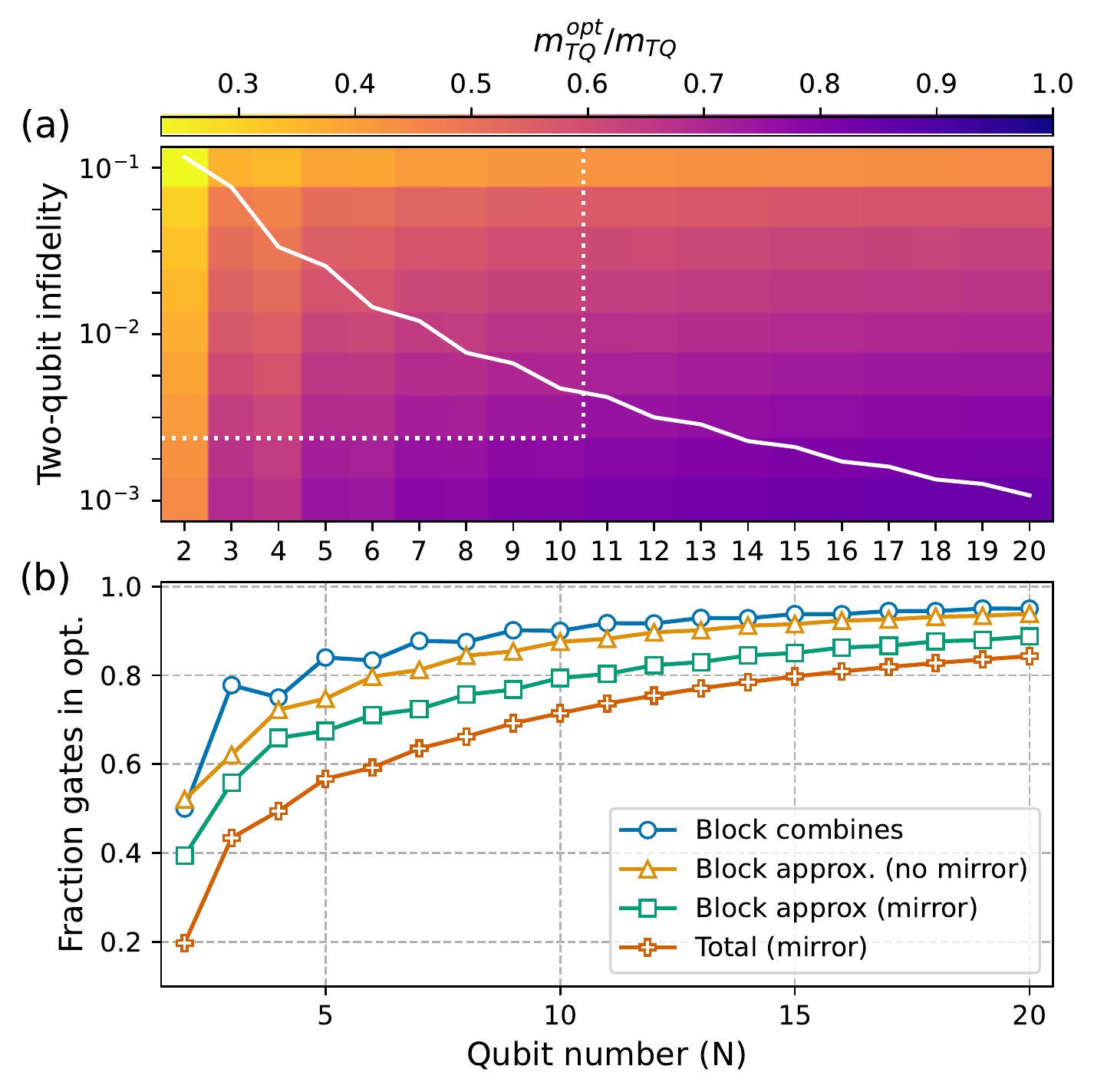}
	\caption{Fraction of two-qubit gates with optimization to without optimization with using block combines (discussed in Sec.~\ref{sec:gate_combine}) and block approximations with mirroring (discussed in Sec.~\ref{sec:gate_approx}). (a) Contour plot of fraction of gates when using both optimizations compared to no optimizations $m_{\textrm{TQ}}(N)/m_{\textrm{TQ}}(N, f)^{\textrm{opt}}$ and plotted as a function of two-qubit infidelity (y-axis) and qubit number (x-axis). The white dashed box outlines the range of QVT$_N$ demonstrations as of writing (QVT$_{10}$ with two-qubit infidelity $\approx 3.16\times 10^{-3}$). The white solid line shows a rough estimate of passing infidelity for each $N$ based on scalable method discussed later in Sec.~\ref{sec:scalable}. (b) Fraction of two-qubit gates of different optimization methods compared to no optimization at estimated two-qubit infidelity required to pass QVT$_N$ from white line in (a) as a function of qubit number.}
	\label{fig:optimizations}
\end{figure}

The combined effect of the block combinations and approximations are plotted in Fig.~\ref{fig:optimizations} over near-term two-qubit infidelity and qubit number ranges. The number of two-qubit gates in a QVT$_N$ circuit is $m_{\textrm{TQ}}(N) = 3 \floor{N/2}N$ without optimizations. Let $m_{\textrm{TQ}}^{\textrm{opt}}(N, f)$ be the average number of two-qubit gates with block combinations and approximations with fidelity threshold $f$.  Fig.~\ref{fig:optimizations}a shows a contour plot where the color gives the ratio between optimized and unoptimized two-qubit gate counts $m_{\textrm{TQ}}(N)/m_{\textrm{TQ}}^{\textrm{opt}}(N, f)$. For small qubit numbers the ratio is small since the block combinations are effective. For large infidelity the ratio is small since the block approximations are effective. The white dashed box outlines the current range of QVT$_N$ realizations as of writing. In the bottom right corner of this range the optimizations reduce the two-qubit gate count to 74\% of the full circuit construction (for $N=10$ and two-qubit infidelity $\approx 3.16\times 10^{-3}$). For larger $N$ the savings from both methods will necessarily decrease since passing will require lower gate infidelity (moving towards the lower right of the figure). 

In Sec.~\ref{sec:numerics} we construct a scalable method to estimate the required infidelity to pass QVT$_N$ for larger $N$. The solid white line in Fig.~\ref{fig:optimizations}a shows the estimated two-qubit infidelity necessary to pass under a two-qubit depolarizing error model. The fraction of two-qubit gates saved at this gate infidelity is plotted as a function of qubit number in Fig.~\ref{fig:optimizations}b to show the relative savings between the methods. The optimizations are very effective at reducing the total number of two-qubit gates for small $N$ (around 50\% reduction for $N=4$) but have diminishing returns as $N$ increases (around 15\% reduction for $N=20$). 

\subsection{Arbitrary angle rotations} \label{sec:arbZZ}
In this section we propose generating SU(4) blocks for QVT circuits using arbitrary-angle two-qubit gates and show that this reduces the error rates per SU(4) block. Achieving arbitrary angle interactions is not a simple task in current experiments but has a variety of applications like in the variational quantum eigensolver (VQE)~\cite{McClean16}, the quantum approximate optimization algorithm (QAOA)~\cite{Farhi14}, and important subroutines like the quantum Fourier transform (QFT)~\cite{Coppersmith02}. 

Enabling arbitrary angle two-qubit interactions, for example $V(\theta) = \textrm{exp}[-i \theta XX/2]$, can potentially reduce the errors per SU(4) block when errors scale with $\theta$. Many two-qubit gate errors scale with $\theta$ such as spontaneous emission in trapped-ion systems or multiplicative rotation errors. As shown below, decomposing SU(4) blocks into arbitrary angle gates reduces the total two-qubit rotation angle per block, and therefore the impact of these types of errors. 

Each SU(4) block in a QVT circuit is decomposed with a Cartan decomposition~\cite{Khaneja01}, which consists of a central two-qubit interaction and single-qubit gates,
\begin{align} \label{eq:arb_angles}
	U &= (K_1 \otimes K_2) V (K_3 \otimes K_4) \nonumber \\
 	V &= \textrm{exp}[-i \tfrac{1}{2}(\theta_x XX + \theta_y YY + \theta_z ZZ)]
\end{align}
where $U\in$SU(4) and $K_i$ are single-qubit gates applied individually to each qubit. Previous tests followed the standard procedure to decompose the middle term into three CNOT gates (or another perfect entangling gate)~\cite{Vatan04}. 

Let $R_{XX}(\theta) = \textrm{exp}[-i \theta XX/2]$ be the available arbitrary angle gate. This is the standard M{\o}lmer-S{\o}renson interaction in trapped-ion experiments with a variable amplitude to change the angle $\theta$~\cite{Sorensen00}. We can rotate $R_{XX}(\theta)$ with single qubit gates to generate $\textrm{exp}[-i \theta YY/2]$ and $\textrm{exp}[-i \theta ZZ/2]$. Since $XX$, $YY$, and $ZZ$ all commute then the middle term can be constructed with three independent applications of $R_{XX}(\theta)$ interleaved with the appropriate single-qubit gates.

\begin{figure} 
	\centering
 	 \includegraphics[width=\columnwidth]{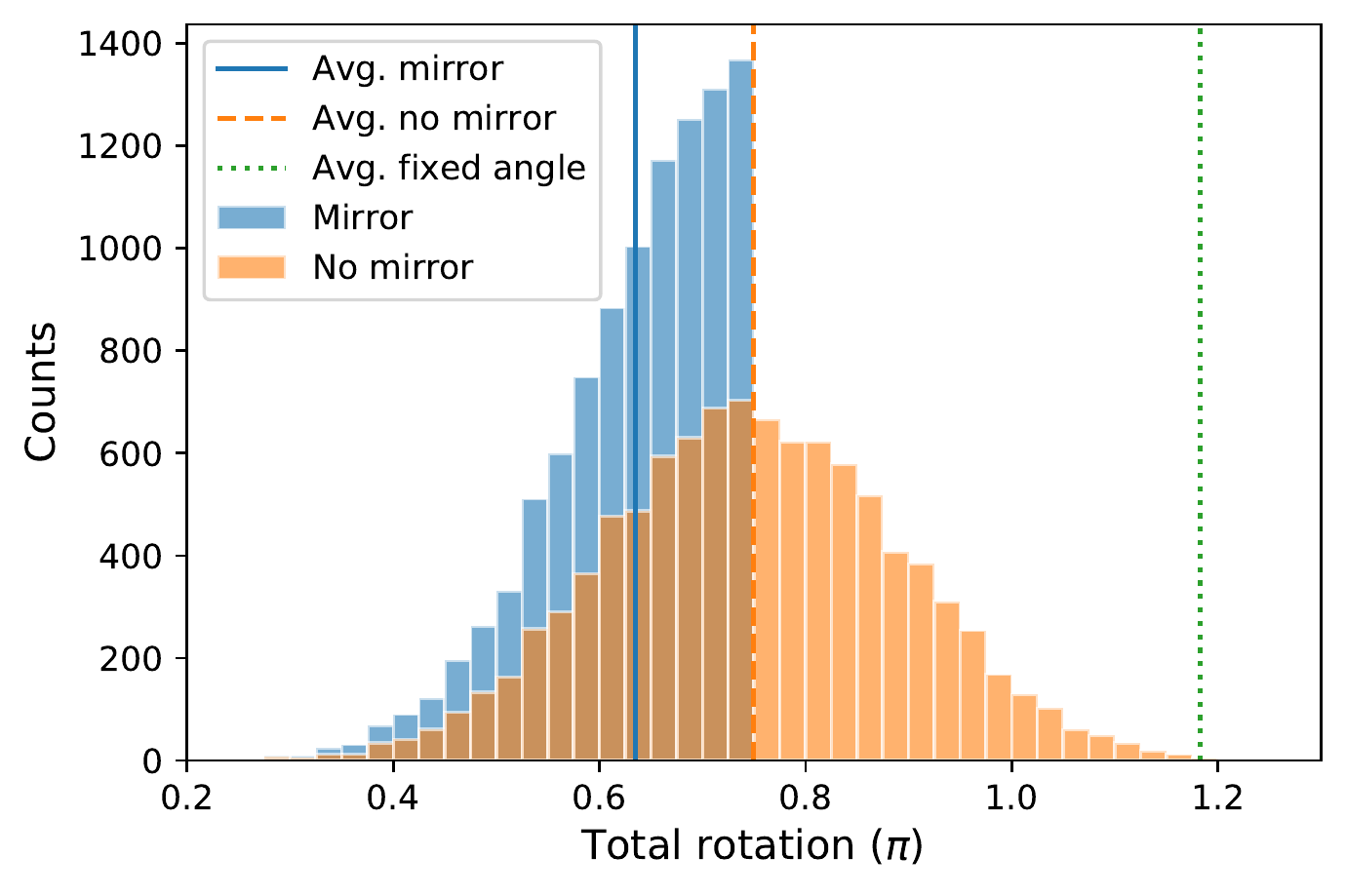}
 	 \caption{Distribution of total rotation angle for arbitrary angle decomposition of 10,000 random SU(4) blocks. Orange histogram shows distribution of total angle. Blue histogram shows the distribution with mirroring option and selecting the decomposition with smallest total angle. Vertical lines show respective means. Green finely dashed line is the average total angle for the approximate method with fixed angle gates and $5\times 10^{-3}$ infidelity.}
\label{fig:arb angle approx}
\end{figure}

We numerically generated 10,000 random SU(4) unitaries with \texttt{Qiskit}~\cite{qiskit} and used its \texttt{TwoQubitWeylDecomposition}, which takes any two-qubit unitary and outputs a standard form based on the Cartan decomposition. For each decomposition we calculated the total rotation angle $\theta_{\textrm{tot}} = |\theta_x| + |\theta_y| + |\theta_z|$ and the distribution is plotted as the orange histogram in Fig.~\ref{fig:arb angle approx}. We also applied  mirroring, which checks the Cartan decomposition to SWAP$\times U$ as described in the previous section, and selected the decomposition that has the smallest total angle. This distribution is the blue histogram in Fig.~\ref{fig:arb angle approx}. For comparison we also plotted the average total angle for the block approximation method as the green finely dashed line with $5 \times 10^{-3}$ infidelity and mirroring.

The arbitrary angle decomposition has less than or equal to the total rotation angle $\theta_{\textrm{tot}}$ of the standard decomposition without block approximations. We empirically observe that using arbitrary angles has average $\theta_{\textrm{tot}} = 3 \pi/4$ (orange dashed line in Fig.~\ref{fig:arb angle approx}) with max $\theta_{\textrm{tot}}=3 \pi/2$. This could be formalized with geometric arguments as in Ref.~\cite{Zhang03}. The mirror option further reduces the total rotation angle with average of $~0.635 \pi$ (blue solid line in Fig.~\ref{fig:arb angle approx}) and the maximum total angle is $3 \pi/4$. The standard decomposition, which consists of three CNOT gates, is equivalent (up to local single-qubit gates) to three applications of $R_{XX}(\pi/2)$, and therefore has $\theta_{\textrm{tot}} = 3 \pi/2$. The arbitrary angle decomposition also has significantly less total rotation angle than the standard method with block approximations as shown with the comparison ot the green finely dashed line of average $\theta_{\textrm{max}} \approx 1.18 \pi$.

One advantage arbitrary angles have over the other optimizations is that the error reduction is constant in qubit number and fidelity. The total rotation angle will be cut in half for any QVT$_N$. With better knowledge of the limiting errors in the arbitrary angle gates further improvements might also be possible.
 
\subsection{Conclusions on circuit constructions}
QVT circuit constructions and optimizations display different behavior for $N<10$ then for $N\geq10$ as well as different behavior for even vs. odd $N$. The ideal success varies 1-2\% with $N$ for $N<10$ (higher values for odd $N$) before approaching the asymptotic value as shown in Fig.~\ref{fig:ideal distribution}. The circuit optimizations also can have significant impact for $N<10$ reducing gate counts between 50\% ($N=4$) and 26\% ($N=10$) in Fig.~\ref{fig:optimizations}. These features imply that running QVT$_N$ for $N>10$ could be more challenging than the previous QVT$_N$ measurements with $N<10$.

\section{Simulating Errors} \label{sec:numerics}
In this section, we present simulations of $\textrm{QVT}_N$ with select error models. We consider errors on three different components: single-qubit gates, two-qubit gates, and measurement. We also consider two other errors that may be missed by component-level benchmarks: memory errors and two-qubit gate crosstalk. Additionally, we propose and test a scalable method for estimating $\textrm{QVT}_N$ success and compare with full numerical simulation for $N<10$. From the scalable method, we estimate error magnitude requirements for $\textrm{QVT}_N$ for various $N$ and error models. We only consider all-to-all connectivity and any extra connectivity constraints almost certainly degrade the performance but are unique to individual systems and compilers. Example implementations of both methods are available on GitHub~\cite{github}.

\subsection{Numerical optimization method}
To simulate $\textrm{QVT}_N$ experiments, we used \texttt{Qiskit}~\cite{qiskit} to generate 5,000 $\textrm{QVT}_N$ circuits for $N=\{2,\dots,9\}$. First, we determine the ideal distribution (without noise) for each circuit with \texttt{Qiskit}'s statevector simulator. Next, each circuit is optimized with a set of custom \texttt{Qiskit} transpiler passes,
\begin{itemize}
	\item \textit{Low:} Combine adjacent single-qubit gates only
	\item \textit{Medium:} $Low$ and SU(4) block combines outlined in Sec.~\ref{sec:gate_combine},
	\item \textit{High:} $Medium$ and block approximation outlined in Sec.~\ref{sec:gate_approx} with mirroring. Fidelity tolerance is selected based on the specified error magnitude, which is the best-case scenario.
\end{itemize}
These passes differ slightly from \texttt{Qiskit}'s built-in transpiler options, which as of writing do not include block approximations, but represent three levels of optimization used in QVT experiments~\cite{Cross19, Pino21, Sundaresan20, Jurcevic21, HQS128, IBM128, HQS512, HQS1024}.

Finally, we apply different noise models with varying magnitudes to each circuit (outlined in Sec.~\ref{sec:error_models} below) and simulate the heavy outputs with a density matrix simulator (\texttt{Qiskit}'s QASM simulator with the snapshot density matrix option), allowing us to distinguish finite sampling effects and circuit noise. The result is a data array of heavy output probabilities with labels [\textit{circuit index}, \textit{qubit number}, \textit{optimization level}, \textit{error model}, \textit{error magnitude}].

\subsection{Scalable estimation method} \label{sec:scalable}
The simulation method outlined above becomes expensive for $N \geq 10$, and here we outline a scalable method based around depolarizing error assumptions and proper accounting of errors. The estimate is constructed by first approximating the fidelity of a single SU(4) block (accounting for the expected number of gates per SU(4) from the transpiling method) and then scaling this estimate for the total number of SU(4) blocks (accounting for the expected number of blocks from the transpiling method). The approach was first proposed in Ref.~\cite{Proctor20} to estimate the success of mirror benchmarking, another full-system benchmark. 

First, we review depolarizing error channels and fidelity. A depolarizing error is a completely-positive trace-preserving (CPTP) quantum map that returns the original state with probability $p$, called the depolarizing parameter, and the maximally mixed state $\mathds{1}/d$ with probability $1-p$~\cite{Nielsen00},
\begin{equation} \label{eq:dep}
	\Lambda[\rho] = p\rho + \tfrac{1-p}{d} \mathds{1}.
\end{equation}
A depolarizing error is simpler to simulate than most errors since it is specified by a single rate and commutes with all operations of the same dimension. We use two fidelity quantities: average fidelity ($F$) and process (or entanglement) fidelity ($f$)~\cite{Nielsen05},
\begin{equation} \label{eq:fidelity}
	\begin{split}
		F(\Lambda) &= \int d \psi \bra{\psi} \Lambda[\ket{\psi}\bra{\psi}] \ket{\psi} =  \frac{(d -1) p + 1}{d},\\
		f(\Lambda) &= \frac{1}{d^2} \textrm{Tr}[\Lambda] =  \frac{(d^2 -1) p + 1}{d^2}.
	\end{split}
\end{equation}
The first equalities in each line are true for all CPTP error channels while the second is specific to a depolarizing error (summarized in Table~1 of Ref.~\cite{CarignanDugas19}). In general, $p \leq f \leq F$ with equality only when $p=1$.

The fidelity of depolarizing errors in a circuit is estimated differently depending on if the corresponding gates are done in parallel or sequentially. A set of parallel gates can be done simultaneously (e.g., in a single round of a QVT circuit across $\floor{N/2}$ pairs). A set of sequential gates must be done in order (e.g., $N$ total rounds in QVT circuits). We can determine how depolarizing errors combine in parallel and sequentially based on the Liouville (or superoperator) representation of quantum processes (reviewed in Refs.~\cite{Greenbaum15, Baldwin20})
\begin{itemize}
	\item \textit{Parallel gates:} Errors on gates performed in parallel on separate qubits have a total process fidelity equal to the product of the individual process fidelities $f_{\textrm{tot}} = \Pi_i f_i$.
	\item \textit{Sequential gates:} $2^N$-dimensional depolarizing errors from gates performed in series on the same $N$ qubits have a total depolarizing parameter equal to the product of the individual depolarizing parameters $p_{\textrm{tot}} = \Pi_i p_i$.
\end{itemize}
The parallel gate relation actually applies to all CPTP processes but the sequential gate relation is specific to depolarizing errors since depolarizing errors commute with the gates in fixed dimension. 

The scalable method works by assuming all errors are depolarizing and that each error can be scaled to cover different numbers of qubits in order to commute errors through gates. For example, pretend the given system has a coherent single qubit error on one qubit. The scalable method treats this error as a single-qubit depolarizing error. If the next gate is a two-qubit gate then that single-qubit depolarizing error will be scaled to be a two-qubit depolarizing error of the same fidelity in order to commute it through the two-qubit gate. Both of these assumptions are unlikely to hold exactly in any experiment or more complicated error model. Errors are never exactly depolarizing channels and errors cannot be scaled in the way we describe below. However, these simplification makes the method scalable to any qubit number and, as shown below, provides reasonable approximations to other, more complicated, error sources.

The first step of the method is to approximate the total error in a single SU(4) block. The SU(4) blocks consist of alternating single-qubit and two-qubit gates (see. Fig.~\ref{fig:gate_combines}a, the final round of single-qubit gates is always combined with the next block). First, we assume all single qubit errors are depolarizing and combine them via the sequential rule above. Next, we determine the process fidelity of two parallel single-qubit gates (each with only single-qubit errors) based on the parallel rule above. Then, we assume that the combined single-qubit processes is a two-qubit depolarizing error and combine it with all the two-qubit errors, which are also assumed to be depolarizing, based on the sequential rule. This produces a net depolarizing rate for each SU(4) block. In principle, other errors like memory or crosstalk can also be combined in this analysis and approximated as depolarizing errors.

The next step is to scale the depolarizing rate per-SU(4)-block to approximate the full circuit error rate. Ref.~\cite{Proctor20} applied the same procedure to combine all blocks of gates at the full-circuit scale. For QVT, as $N$ increases this method is roughly equivalent to raising the \textit{process fidelity} per-SU(4)-block to the $n_{\textrm{rounds}} \floor{N/2}$ power, where $n_{\textrm{rounds}}$ is the number of rounds determined by the transpiler optimization. We find this method mostly underestimates the actual success when compared to numerical simulation.

As an alternative, we raise the \textit{average fidelity} per-SU(4)-block to the power of $n_{\textrm{rounds}} \floor{N/2}$. We find that this method approximates the actual success better in simulations for $N\leq 9$. The reasons why this is a better approximation likely relate to the ways errors spread in QVT circuits but we leave a complete study for future work. 

The resulting estimates from either method is used as an approximation of the depolarizing rate for the entire circuit. This error produces the correct output state with probability $p_{\textrm{tot}}$, which has the ideal success $h_{\textrm{ideal}}(N)$, and the maximally mixed state $\mathds{1}/2^N$ with probability $1 - p_{\textrm{tot}}$, which will return heavy outputs half the time. 

Finally, for both options we include measurement errors, which return a false output with probability $e_M$ per qubit. Therefore, the probability of measuring the correct outputs for the entire circuit is $p_M = (1 - e_M)^N$. We assume that any error in the measurement produces a heavy output half of the time. Later, we use both methods to define an estimated region of QVT$_N$ success.

The method is summarized in Algorithm~\ref{alg:scalable}. We define three functions. First, $\texttt{convert}(x, \textit{avg}\rightarrow\textit{proc})$ converts $x$ between different quantities (e. g. average $\rightarrow$ process fidelity with abbreviations average fidelity = \textit{avg}, process fidelity = \textit{proc}, and depolarizing parameter  = \textit{dep}). Next, $\texttt{rounds}(N)$ returns the number of parallel rounds of SU(4) blocks based on Sec.~\ref{sec:gate_combine}. Finally, $\texttt{gates}(tol)$ returns the number of gates per SU(4) block based on Sec.~\ref{sec:gate_approx} for given $tol$ average fidelity level.

\begin{algorithm}[H]
	\caption{Scalable estimation of QVT$_N$}
	\label{alg:scalable}
	\begin{algorithmic}[1]
		\Procedure{Scalable}{$errors$, $N$, $opt$, $method$, $tol$}
		\If{$opt$ $= \textit{low}$}
		\State $m= 3$  \Comment{Two-qubit gates per SU(4) block}
		\State $n = \floor{N/2} N$  \Comment{Total number of SU(4) blocks}
		\ElsIf{$opt$ = $\textit{medium}$}
		\State $m = 3$
		\State $n =\floor{N/2} \texttt{rounds}(N)$
		\ElsIf{$opt$ = $\textit{high}$}
		\State $m = \texttt{gates}(tol)$
		\State $n = \floor{N/2} \texttt{rounds}(N)$
		\EndIf 
		\State $p_{\textrm{SQ}} = \Pi_i \texttt{convert}(F_{i, \textrm{SQ}}($errors$), \textit{avg}\rightarrow\textit{dep})$ \Comment{Product over all single-qubit error sources}
		\State $p_{\textrm{TQ}} = \Pi_i \texttt{convert}(F_{i, \textrm{TQ}}($errors$), \textit{avg}\rightarrow\textit{dep})$  \Comment{Product over all two-qubit error sources}
		\State $p_{\textrm{SU}(4)} = (p_{SQ} \times p_{TQ})^{m}$
		
		\If{$method$ $= \textit{avg}$}
		\State $p_{\textrm{tot}} =\texttt{convert}(p_{\textrm{SU(4)}}, \textit{dep}\rightarrow\textit{avg})^{n}$ 
		
		\ElsIf{$method$ $= \textit{proc}$}
		\State $p_{\textrm{tot}} =\texttt{convert}(p_{\textrm{SU(4)}}, \textit{dep}\rightarrow\textit{proc})^{n}$ 
		\EndIf 
		\State $p_M = e_M^N$  \Comment{Measurement error}
		\State $s = h_{\textrm{ideal}}(N) p_{\textrm{tot}} p_M + (1 -  p_{\textrm{tot}} p_M)/2$ 
		\State \textbf{return} $s$
		\EndProcedure
	\end{algorithmic}
\end{algorithm}

\subsection{Types of errors} \label{sec:error_channels}
We simulate $\textrm{QVT}_N$ with the following errors.
\begin{itemize}
\item \textit{Single-qubit errors:}  QVT$_N$ circuits contain $7 \floor{N/2} N$ single-qubit gates without optimization, (although $2 \floor{N/2} (N-1)$ are eliminated with all transpiler passes outlined above). We model single-qubit errors as depolarizing as given in Eq.~\ref{eq:dep} with $d=2$.

\item \textit{Two-qubit errors:} QVT$_N$ circuits contain $3 \floor{N/2} N$ two-qubit gates without optimization. We model two types of two-qubit errors: two-qubit depolarizing (as given in Eq.~\ref{eq:dep} with $d=4$) and coherent $ZZ$ rotations (a common error for devices whose native two-qubit gate is based on a $ZZ$ (or a $XX$ or $YY$) interaction~\cite{Pino21}).

\item \textit{Measurement errors:} At the end of each circuit, $N$ single-qubit measurements are made. A measurement error of probability $p_M$ falsely returns a ``1'' (or ``0'') output when the measurement operation actually projected the qubit into ``0'' (or ``1'').  In practice, the two qubit states may have different false measurement output probabilities, but we assume they are equal for simplicity.
\end{itemize}

\begin{table*}[] 
	\centering 
	\resizebox{0.8\paperwidth}{!}{
	\begin{tabular}{lcccccc}
		\hline
		\hline
		Error model         & SQ depolarizing & TQ depolarizing & TQ coherent & TQ memory & TQ crosstalk & Measure \\ \hline
		\textit{SQ depolarizing}     & 10   & 1     & 0                & 0                   & 0                  & 1       \\
		\textit{TQ depolarizing}     & 1  &  10    &     0           &    0                 &    0               & 1        \\
		\textit{TQ coherent}          & 1  &  0      &  10       &    0                &    0                &  1       \\ 
		\textit{Measurement}   &  1 &  10   &  0              &    0                &   0                 &  10     \\
		\textit{Crosstalk} &  1 &  10   & 0               &     0              &   1                 &   1      \\
		\textit{Memory}   &  1  &  10  & 0               & 1                  &  0                  & 1       \\
		\textit{TQ mixed}          &  1  & 5    &  5               &    0               &  0                  & 1     \\
		\textit{Semi-realistic}   & 1  & 10  & 1                 & 1/2                 &  1/2                  & 1     \\ 
		\hline
		\hline
	\end{tabular}
	}
	\caption{Error models with names given in first column based on dominant sources of errors. Each error model has a different ratio of error sources (named in first row) that are combined to determine a realization based on an error magnitude $\varepsilon$.} \label{table:error_models}
\end{table*}

We also model two common types of full system errors:
\begin{itemize}
\item \textit{Memory errors:} There are several instances of idle qubits in QVT circuits where memory errors can occur. First, for odd $N$ a single qubit will be left out of each gate round. Second, qubits may be left idle if gates are not able to be performed in parallel either by design, such as in Ref~\cite{Pino21}, or to avoid crosstalk errors as in Ref.~\cite{Jurcevic21}. Here, we add single-qubit dephasing errors before every two-qubit gate as a simple example of memory errors.

\item \textit{Crosstalk errors:} Crosstalk errors usually refer to unintended operations on qubits caused by nearby gates~\cite{Sarovar20}, and are architecture dependent. Assuming a linear array of qubits, we model crosstalk errors caused by two-qubit gates as single-qubit depolarizing errors on nearest neighbor qubits. 
\end{itemize}

\subsection{Error models} \label{sec:error_models}
We ran numerical simulations with several different error models to examine the sensitivity of QVT$_N$ to commonly structured noise environments. Each error model is specified by scaling factors for the various error sources introduced in Sec.~\ref{sec:error_channels}, and the models are defined in Table~\ref{table:error_models}. Error models are written with script font to differentiate from error sources and the abbreviations single-qubit (SQ) and two-qubit (TQ) are used for brevity. The first four models highlight different component errors (\textit{SQ depolarizing}, \textit{TQ depolarizing}, \textit{TQ Coherent}, and \textit{Measurement} models). The next four models contain some level of several errors to better approximate real systems where multiple errors are present at different magnitudes but different sources dominate (\textit{Crosstalk}, \textit{Memory}, \textit{TQ mixed}, and \textit{Semi-realistic} models). 

A realization of a given error model is determined by a single error magnitude $\varepsilon$. This magnitude is scaled by the values in Table~\ref{table:error_models} to determine the average infidelity of each error source. For measurement errors the scaled error magnitude is equal to the probability of returning the incorrect output. We ran simulations with seven different error magnitudes for each error model exponentially distributed between $[10^{-3.25}, 10^{-1.25}]$.

The SQ depolarizing, TQ depolarizing, and TQ coherent error sources were normalized such that the estimated infidelity of a block of two single- and one two-qubit gate is equal to the specified error magnitude (discussed further below). The \textit{SQ depolarizing}, \textit{TQ depolarizing}, and \textit{TQ coherent} models all have the same estimated infidelity per single- and two-qubit block equal to the error magnitude. This facilitates direct comparisons between different error models that produce similar estimates of fidelity in component level experiments like randomized benchmarking. For example, coherent errors and depolarizing errors lead to similar fidelity estimates in randomized benchmarking experiments but coherent errors may be more detrimental to other quantum circuits~\cite{CarignanDugas19}. Memory and crosstalk errors are excluded from this normalization since these errors may be missed by a randomized benchmarking experiment. Measurement errors are also excluded since they are measured separately.

For the normalized error sources, the normalization constant $n$ is found from the early steps in Algorithm~\ref{alg:scalable} under a small error approximation. The approximated average fidelity of a two single- and one two-qubit gate block is set equal to the error magnitude with normalization,
\begin{equation}
	\varepsilon = \frac{1}{n} \left[\tfrac{12}{5} \sum_i  s_i (1 - F_{SQ, i}) + \sum_j s_j (1 - F_{TQ, j})\right],
	\label{eq:tot_fid}
\end{equation}
where $i$ labels single-qubit error sources with average fidelity $F_{SQ, i}$ and $j$ labels all two-qubit error sources with average fidelity $F_{TQ, j}$. The scaling factors $s_i$ and $s_j$ are the constants in Table~\ref{table:error_models}. This allows us to solve for $n$ given $s_{i/j}$ when $F_{SQ, i} = F_{TQ, j} = 1-\varepsilon = 0$. For example, with the \textit{SQ depolarizing} model $n = 25$. The parameters used to define each error source are generated by scaling the error magnitude by $s_{i/j}/n$. For the error sources considered,
\begin{equation}
	\begin{split}
		p_{\textrm{SQ}} &= 2 s_{0} \varepsilon/n, \\
		p_{\textrm{TQ}} &= 4 s_{1} \varepsilon/3 n,\\
		\theta_{\textrm{TQ}} &= 2 \arccos\sqrt{\tfrac{4 - 5 s_{2} \varepsilon/n}{4}} ,
	\end{split}
\end{equation}
where $p_{\textrm{SQ}}$ and $p_{\textrm{TQ}}$ are the depolarizing single- and two-qubit rates respectively,  $\theta_{\textrm{TQ}}$ is the rotation angle for two-qubit coherent errors, and the scaling parameters are indexed in the same order.

\begin{figure*} 
	\centering
		\includegraphics[width=0.94\textwidth]{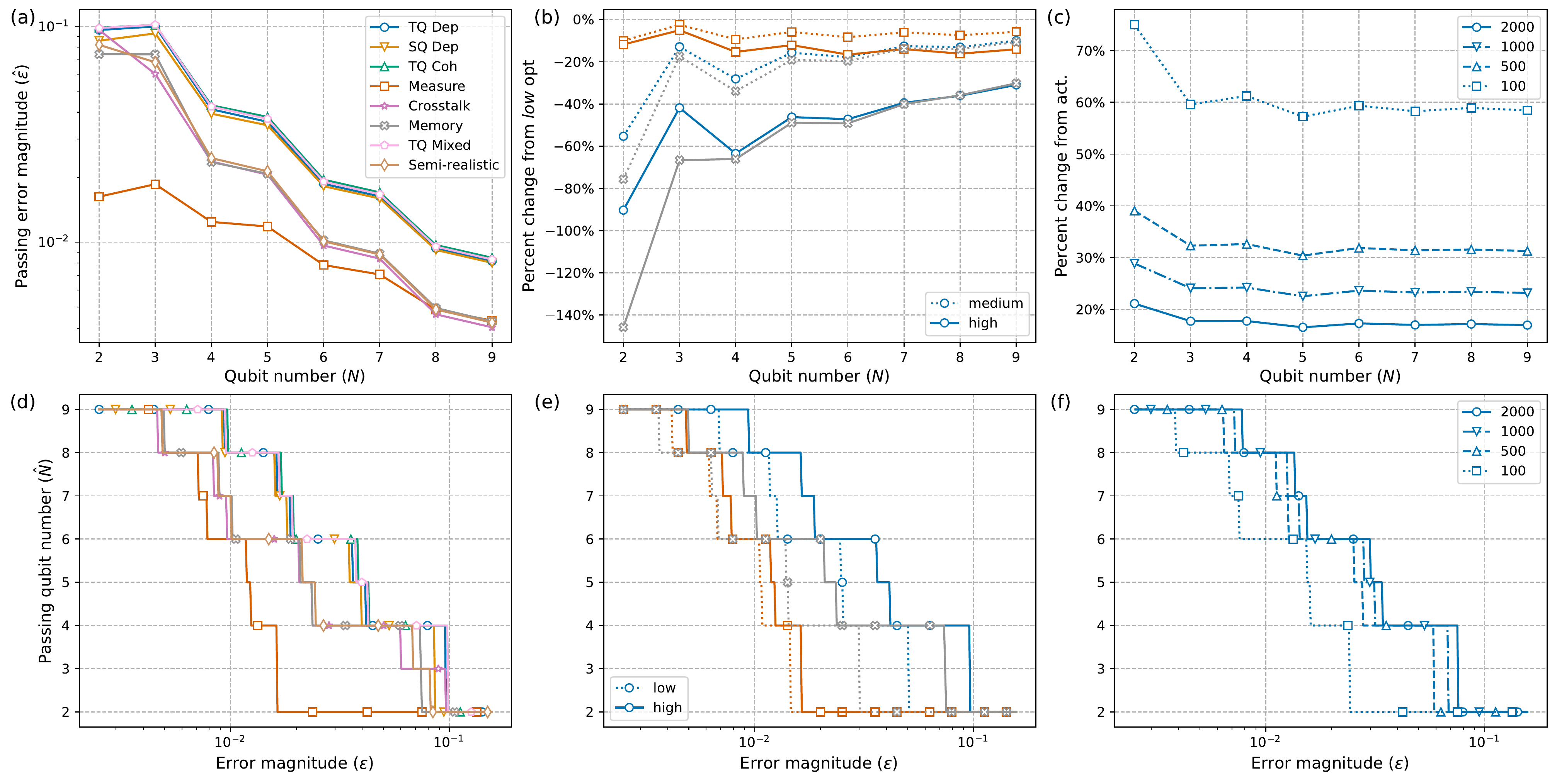}
		\caption{Numerical passing threshold estimates from interpolation of QVT$_N$ for $N = \{2, \dots, 9\}$ with various optimization levels, error models and magnitudes. (a) Passing error magnitude $\hat{\varepsilon}$ (error magnitude when estimated success equals 2/3) for QVT$_N$ and $high$ optimization for each error model as a function of $N$. (b) Percent change in passing error magnitude for three example error models (\textit{TQ depolarizing}, \textit{Measure}, and \textit{Memory} with same colors as a). Line styles indicate different optimization methods. (c) Percent change in passing error magnitudes for \textit{TQ depolarizing} model and $high$ optimization for different numbers of circuits. Line styles indicate different number of circuits. (d) Passing qubit number $\hat{N}$ as a function of error magnitude (estimated success is above 2/3) for each error model and $high$ optimization. (e) Passing qubit number as a function of error magnitude for different optimization levels and error model (\textit{TQ depolarizing}, \textit{Measurement}, and \textit{Memory} with same colors as b). (f) Passing qubit number as a function of error magnitude for different number of total circuits with \textit{TQ depolarizing} model and $high$ optimization.}
		\label{fig:numeric_data}
\end{figure*}

\subsection{Numerical results}
Here, we present simulations of the QVT$_N$ solving for the passing requirements for different qubit numbers, error models, optimization levels, and circuit samplings. To this end, we generated 5,000 random QVT$_N$ circuits for $N = 2 - 9$ and estimated the success for transpiler optimization methods $low$, $medium$, and $high$, eight different error models (Table~\ref{table:error_models}), and seven exponentially distributed error magnitudes ($\varepsilon \in [10^{-3.25}, 10^{-1.25}]$), for a total of 1,512 different settings and 7,560,000 simulated circuits. For $N=2 - 6$ we ran an additional dataset with larger $\varepsilon$ to sample success rates below 2/3. For a given qubit number, error model, magnitude and optimization level we assume the simulated average heavy output probability with errors over all 5,000 circuits is approximately equal to the success (average over $all$ QVT circuits).

We present the data in terms of the minimum requirements to pass the QVT$_N$ test from both a qubit limited and a fidelity limited perspective. If the system is \textit{qubit limited} the main question is what fidelity is required to pass QVT$_N$ for a given qubit number? This perspective is plotted in Fig.~\ref{fig:numeric_data}a-c. If the system is \textit{fidelity limited} the main question is what qubit number $N$ can pass QVT$_N$ with a given fidelity? This perspective is plotted in Fig.~\ref{fig:numeric_data}d-e. Below, we mostly follow the qubit limited perspective but translate all results to the fidelity limited view in parentheses. 

Fig.~\ref{fig:numeric_data}a (and d) plots the estimated maximum error magnitude $\hat{\varepsilon}$ that passes QVT$_N$ for each error model as a function of qubit number $N$ (and estimated maximum passing qubit number $\hat{N}$ as a function of error magnitude). We refer to this maximum as the passing threshold $\hat{\varepsilon}$ (or qubit number $\hat{N}$) where the estimated success is equal to 2/3 determined by cubic spline interpolation of the dataset with fixed $N$ (and interpolation of both $N$ and $\varepsilon$). The models that are dominated by gate errors, e.g. the \textit{TQ depolarizing}, \textit{SQ depolarizing}, \textit{TQ coherent}, and \textit{TQ mixed} models, have the easiest to achieve passing thresholds. When we add extra errors beyond the standard component-level error sources, e.g. in the \textit{Crosstalk}, \textit{Memory}, and \textit{Semi-realistic} models, the test is harder to pass and the estimated passing threshold. This matches with experimental results where performance on system-level benchmarks is often worse than predicted from component-level benchmarks~\cite{Lubinski21, Proctor20} since often there are unmeasured errors present.

Overall, the estimated passing thresholds fall into three groups based on the error model's total magnitude and not the type of errors. First, \textit{SQ depolarizing}, \textit{TQ depolarizing}, \textit{TQ coherent}, and \textit{TQ mixed} all have the same magnitude per single- and two-qubit gate round (as defined in Sec.~\ref{sec:error_models}) but different types of errors dominate. Second, \textit{Crosstalk}, \textit{Memory}, and \textit{Semi-realistic} all have similar magnitude per single- and two-qubit gate round but they also include other error sources with similar magnitudes. Finally, \textit{Measurement} has a different scaling with $N$ since measurement errors dominate but there are a linear number of measurements versus a quadratic number of gates. However, this difference is only observable for small $N$. As $N$ increases the \textit{Measurement} model begins to scale more similarly to other models since the number of measurements is much smaller than the number of gates. The similarities within each group may be partially due to the types of errors we selected but also implies that the QVT$_N$ is mostly sensitive to total error magnitude and not type of error or other metrics like diamond norm~\cite{Nielsen05}. This is not wholly unexpected for random circuit averaging and is seen in similar methods like randomized benchmarking~\cite{CarignanDugas19}. 

Fig.~\ref{fig:numeric_data}b (and e) shows the effectiveness of the different transpiler options: the $medium$ (block combines from Sec.~\ref{sec:gate_combine}) and $high$ (block combines and approximations from Sec.~\ref{sec:gate_approx}). The passing threshold is again estimated from interpolation for the $high$ (solid lines), $medium$ (dashed lines), and the $low$ optimization (finely dashed in e). We plot three example models (\textit{TQ depolarizing}, \textit{Measure}, and \textit{Memory}) that represent the three different groups of error models seen in Fig.~\ref{fig:numeric_data}a (and d), showing the reduced effectiveness as qubit number increases, as expected based on Sec.~\ref{sec:optimizations}. For the \textrm{Measurement} model, the optimization methods considered are not as effective since these methods are not aimed at measurement errors but other mitigation methods may be more effective~\cite{Maciejewski20}.

For any optimization and error model, a realization of a QVT$_N$ experiment will always require lower error magnitude (or only pass for lower qubit number) than what is plotted in Fig.~\ref{fig:numeric_data}a (and d) due to the confidence interval requirement to pass QVT$_N$. This means the success must clear a higher threshold than $2/3$, which is dependent on the number of circuits run. Fig.~\ref{fig:numeric_data}c (and f) show the percent change in error magnitude (and passable $N$) for different total number of circuits extracted from cubic spline interpolation. By the definition in Ref.~\cite{Cross19}, the confidence interval is independent of $N$, and therefore the percent change is proportional to the square-root of the number of circuits. For $N=2$ the ideal success is much lower, which also changes the confidence interval based on Eq.~\eqref{eq:confidence_ibm}.

\begin{figure} 
	\begin{center}
	  \includegraphics[width=\columnwidth]{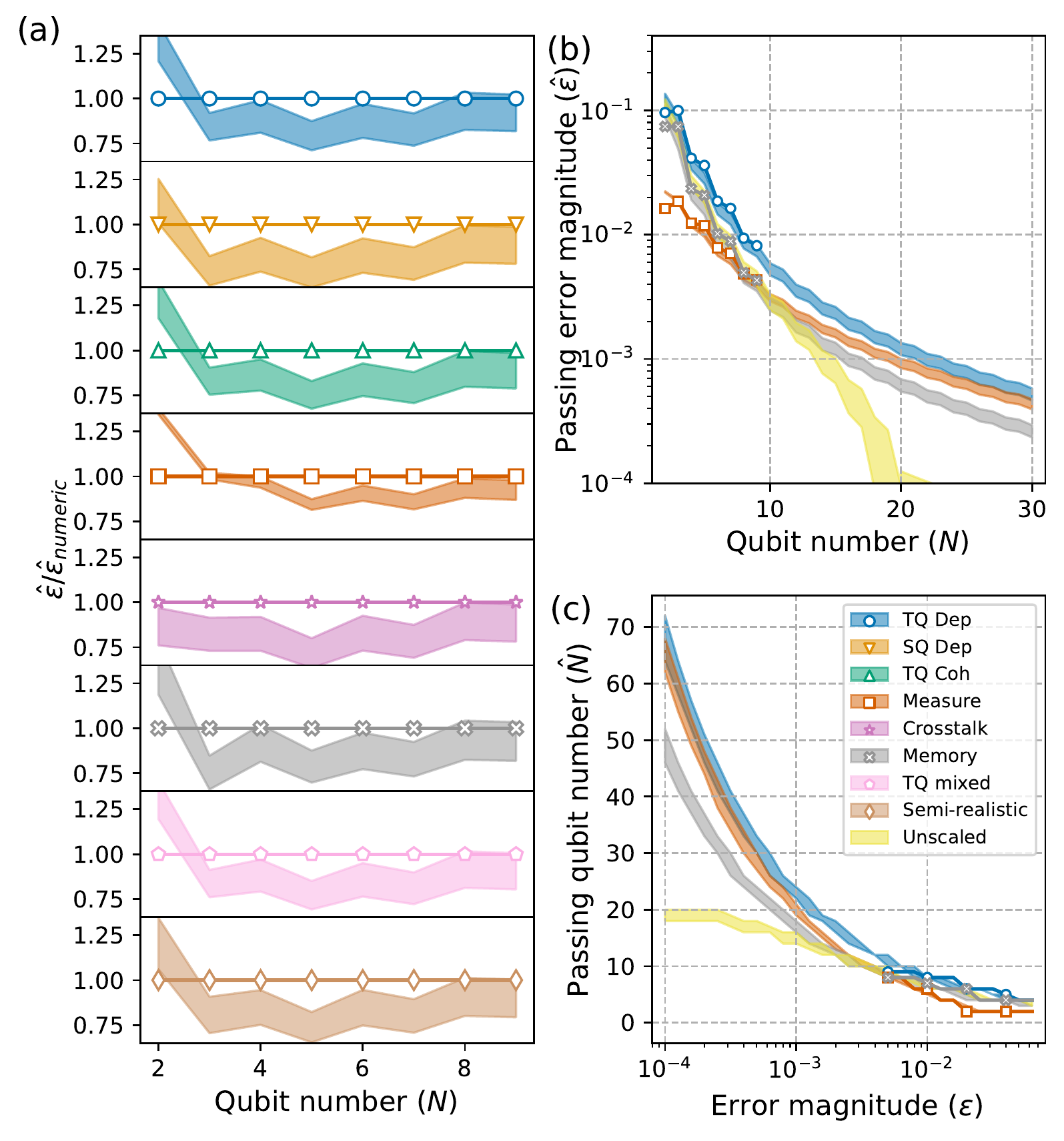}
	  \caption{Comparisons between numerical simulations and the scalable method.  (a) Ratio between the passing error magnitude found from different methods and the passing error magnitude estimated from numerical simulation $\hat{\varepsilon}_{\textrm{numeric}}$. Colored regions correspond to the scalable method bounded between $avg$ and $dep$ options. Solid line is ratio=1 to compare how closely the scalable method matches the numerical results.(b) Qubit limited scaling of passing error magnitude $\hat{\varepsilon}$ vs qubit number $N$ with full simulation data (solid lines and circles) compared to the scalable method (colored regions filled between $dep$ and $avg$ options). We added an additional model called $Unscaled$ (yellow region) that has a fixed magnitude crosstalk error of $10^{-3}$. (c) Fidelity limited scaling of passing qubit number $\hat{N}$ vs error magnitude with full simulation data (solid lines and circles) compared to the scalable method.}
	\label{fig:numeric_comparisons}
	\end{center}
\end{figure}

Next, we study the scalable method's effectiveness at predicting the passing threshold, summarized in Fig.~\ref{fig:numeric_comparisons}. In Fig.~\ref{fig:numeric_comparisons}a we compare the predicted passing error magnitude from the scalable method (colored regions) to the estimation from full simulation data (points and lines) for each error model. We find that both scalable methods are within 25\% difference of the simulated data for $N < 10$ but underestimate the required error magnitude (predicts error magnitudes that are harder to achieve). However, both scalable methods mostly overestimate the required error magnitude for $N=2$ (predicts error magnitudes that are easier to achieve). This is not a large impediment since most systems are far beyond QVT$_2$. 

In Fig.~\ref{fig:numeric_comparisons}b and c we use the scalable model (colored regions) to make predictions about QVT$_N$ requirements for $10 \leq N \leq 30$ in the (b) qubit limited and (c) fidelity limited perspectives for three example error models: \textit{TQ depolarizing}, \textit{Memory}, and \textit{Measurement}. In both plots we added an additional error model called $Unscaled$ that is the \textit{TQ depolarizing} model with an additional fixed magnitude crosstalk error of $10^{-3}$. The three original error models perform as expected: more errors increase the requirements of the error magnitude so we expect \textit{Memory} to require lower errors than \textit{TQ depolarizing}. For \textit{Measurement} the measurement errors are an order of magnitude larger than two-qubit errors. For small $N$ the required error magnitude scales with the number of measurements $N$ but with larger $N$ there are many more two-qubit gates that cause the error magnitude to scale with $N^2$ and the performance approaches the \textit{TQ depolarizing} model. The \textit{Unscaled} model has an error that cannot be lowered, and therefore sets a hard limit for QVT$_N$ of $N\approx20$. This also affects the requirements for $N<20$ as seen by the divergence between the \textit{Unscaled} and \textit{TQ depolarizing}.

Fig.~\ref{fig:numeric_comparisons}b and c can also be used to make predictions for fidelity needed to demonstrate quantum computational advantage in sampling. For instance, take $N=50$ as a possible point that QVT circuits will no longer be simulatable. The scalable method predicts that the \textit{TQ Depolarizing} model requires two-qubit gate fidelity to be $~2\times10^{-4}$ to pass QVT$_{50}$. However, QVT circuits might not be the most efficient method for such a demonstration, e.g. Ref.~\cite{Arute19} uses less gates and lower fidelity.

\begin{figure*} 
	\begin{center}
		\includegraphics[width=\textwidth]{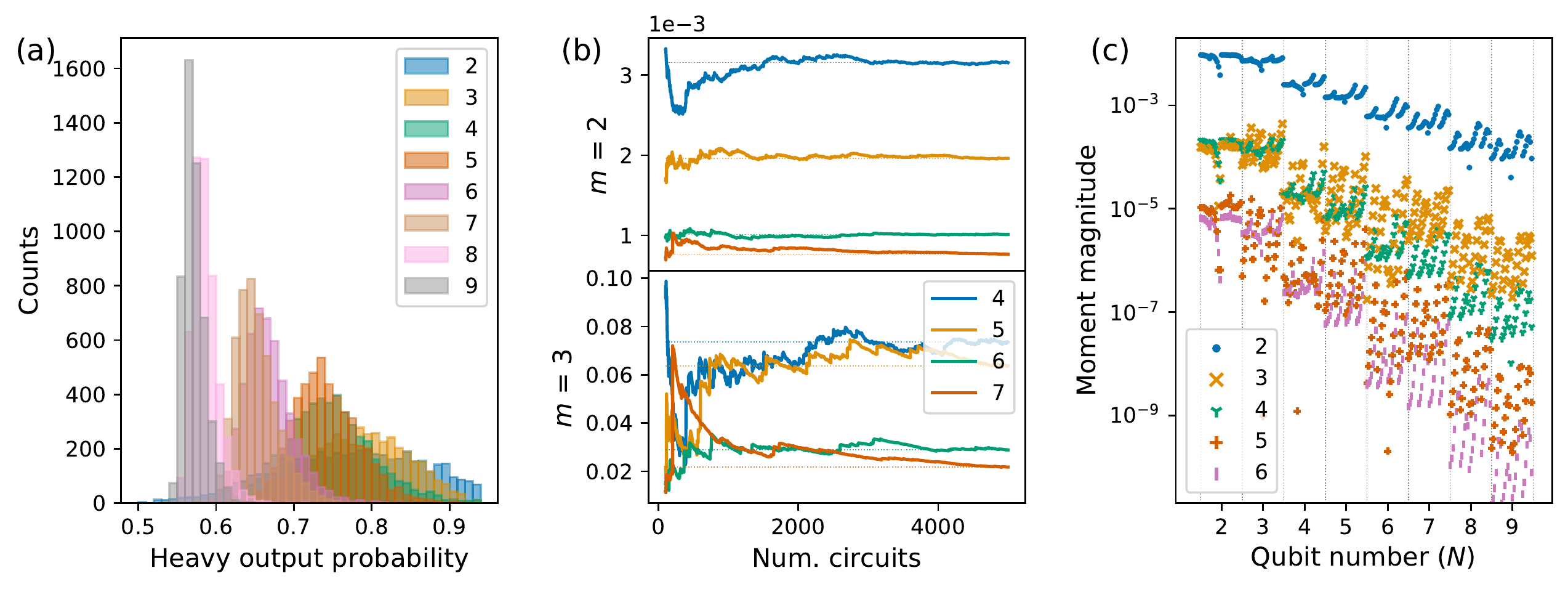}
		\caption{Numerical study of heavy output distributions all with $high$ optimization. (a) Example histograms for various $N$ from the \textit{Semi-realistic} error model with $\varepsilon=0.01$. (b) Numerically estimated second and third moments ($m$) for $N=4, 5, 6, 7$ as a function of number of circuits. (c) Numerically estimated moments ($m=2,3,4,5,6$) from \textit{TQ depolarizing}, \textit{Measurement}, \textrm{TQ mixed}, \textrm{Semi-realistic} error models, all sampled magnitudes, and all qubit number flattened into a single index on the x-axis. Horizontal divisions show grouping of different qubit numbers.}
		\label{fig:distributions1}
	\end{center}
\end{figure*}

\subsection{Conclusions on simulations}
Based on our limited simulations, the passing threshold for QVT$_N$ is more dependent on total error magnitude than the type of error as seen in the different error models in Fig.~\ref{fig:numeric_data}. Moreover, the good agreement between full numerical simulations and scalable approximate simulations in Fig.~\ref{fig:numeric_comparisons} shows that our method does a decent job of capturing the scaling of the required error magnitude to pass QVT$_N$ but mostly returns conservative estimates. This leaves room for improvement and open questions about how errors are spread in QVT and other circuits.

\section{Confidence intervals} \label{sec:confidence}
The confidence interval lower bound defines the passing criteria for QVT$_N$. As previously defined, let $\hat{h}$ be the average heavy output frequency of a set of measured circuits with finite sampling statistics and let $h$ be the average heavy output probability with errors over \textit{all} $\textrm{QVT}_N$ circuits (success). A two-sigma confidence interval certifies that the confidence interval computed from the measured data contains $h$ 97.73\% of the time. 

In Ref.~\cite{Cross19}, the confidence interval is constructed assuming that each circuit is run with a single shot, the measured heavy output frequency is then a binomial random variable with probability $\hat{h}$, and it is assumed there are enough circuits that the distribution is Gaussian (defined as at least 100). This was viewed as a conservative approach since in most experiments more than one shot is run per circuit. The confidence interval is estimated based on the binomial variance and solely on the total number of random circuits (not the shots per circuit). With an equal number of shots per circuit the confidence interval is,
\begin{equation}
C_{\textrm{lower}} = \hat{h} - 2 \sqrt{\frac{\hat{h} (1 -\hat{h})}{n_c}}.
\end{equation}

One problem with this confidence interval estimate is that the average heavy output frequency is usually calculated from more than one shot per circuit and so not necessarily binomial. The heavy output frequency from an individual circuit is a binomial random variable with probability $h_i$, but that probability $h_i$ has a distribution determined by the initial circuit, optimizations, and noise environment. An example of this distribution is shown in Fig.~\ref{fig:distributions1}a for various $N$ with a fixed error model and magnitude constructed from a sample of 5,000 QVT$_N$ circuits. Fig.~\ref{fig:distributions1}b shows that the $m=2$ (variance) and $m=3$ (skewness) moments have mostly stabilized at 5,000 circuits and the values decrease with $N$. Therefore, 5,000 circuits gives a reasonable representation. The average of $h_i$ over sampled circuits, is not binomial, but the variance can be bounded by the binomial sum variance inequality~\cite{Hoeffding56}. 

We propose and test a method to construct tighter confidence intervals that account for this distribution across circuits and still covers 97.73\% of experiments. This method is based on a semi-parametric bootstrap technique originally proposed for randomized benchmarking~\cite{Meier06}. An example implementation of this method and comparison to the original method is available on GitHub~\cite{github}.

Bootstrapping is a technique to construct confidence intervals based on repeatedly sampling from a dataset~\cite{Efron94}. Given a QVT dataset that consists of $n_c$ circuits each with $n_s$ shots. We first sample ``non-parametrically" $n_c$ circuits with replacement (not removing any sampled circuit from the dataset for future sampling). For each of these circuits, we then sample ``parametrically" $n_s$ shots from a binomial distribution with probability equal to the measured heavy output frequency of that given circuit. This whole procedure is repeated many times to produce a distribution of heavy output frequencies. From this distribution we calculate quantiles that cover any percentage of the data and use those quantiles to construct confidence intervals. The method is summarized as follows:
\begin{enumerate}
\item Randomly sample $n_c$ circuits from the dataset with replacement 
\item For each of the $n_c$ circuits, randomly sample $n_s$ shots based on a binomial distribution with the probability set to the heavy output frequency of the given circuit $h_i$
\item Estimate the average sampled heavy output frequency $\hat{r}$
\item Repeat steps (1-3) $n_b$ times to form the distribution $\{ \hat{r}_i \}$
\item Calculate the lower two-sigma confidence interval based on the distribution $\{ \hat{r}_i \}$ and basic bootstrap confidence interval $C_{\textrm{lower}} = 2 \overline{r} - Q(\{ \hat{r}_i \},\, 97.73\%)$~\cite{Efron94} where $\overline{r}$ is the mean of  $\{ \hat{r}_i \}$
\end{enumerate}
Step 1 reflects the statistical fluctuations from randomly sampling QVT circuits, while step 2 reflects the quantum statistical fluctuations from measuring the circuits. The quantile function $Q(\{ \hat{r}_i \}, \, 97.73\%)$ returns a threshold that is greater than 97.73\% of the distribution $\{\hat{r}_i\}$.

To test the confidence intervals we check the coverage probability and interval widths. The coverage probability is the fraction of times the true value is contained within the confidence interval for asymptotically many repetitions of the experiment. In the case of QVT, the coverage probability is the fraction of cases where the actual success (average heavy output probability over all possible QVT circuits) for a given QVT$_N$ and error model is contained within the constructed confidence interval. The interval width is the distance between the measured value and the confidence interval bound. For QVT, we only study the lower width, which is the distance between the measured heavy output frequency and the constructed lower two-sigma confidence interval.

To test the coverage probability and confidence interval widths, we sample from the numerical data generated in Sec.~\ref{sec:numerics} for the \textit{TQ depolarizing}, \textit{Measurement}, \textit{TQ mixed}, and \textit{Semi-realistic} error models. Again, we assume that the average heavy output probability with errors over the 5,000 circuits sample for given qubit number, error model, magnitude and optimization level is approximately equal to the success. In Fig.~\ref{fig:distributions1}b we show that this is a good approximation since the moments of this distribution stabilize as more circuits are simulated. In Fig.~\ref{fig:distributions1}c we study the moments for each distribution for all sets of 5,000 circuits and see that in fact the second through sixths moments shrink mostly with qubit number and some dependence on errors. The plotted data only shows the absolute value of the moments, but some odd number moments are in fact negative for small qubit number, which indicates a small amount of skewness in the distributions.

\begin{figure} 
	\begin{center}
	  \includegraphics[width=\columnwidth]{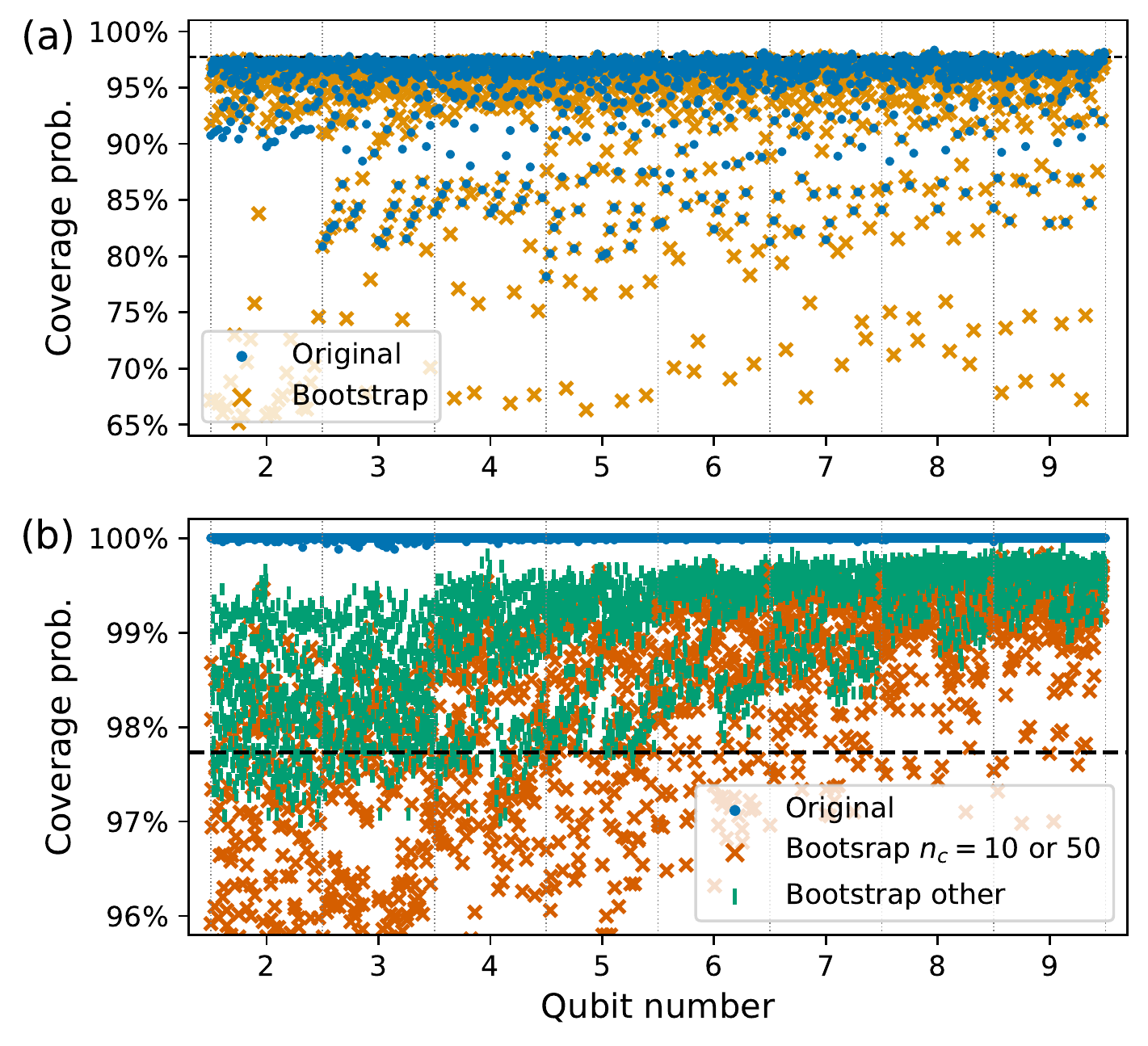}
	  \caption{Comparison of confidence interval methods. (a) Coverage probability for single shot ($n_s=1$) experiments for semi-parametric bootstrap resampling (orange ``x'') compared to original confidence interval proposed by Cross \textit{et al.}~\cite{Cross19} (blue points). Most points are below the 97.73\% expected coverage indicating the methods are not constructing proper confidence intervals for single shot experiments. (b)  Coverage probability for all other $n_s$ for semi-parametric bootstrap resampling with less than 100 circuits (red ``x'') and more than 100 circuits (green lines). Original confidence interval construction from by Cross \textit{et al.}~\cite{Cross19} (blue points).}
	\label{fig:distributions2}
	\end{center}
\end{figure}

For a given qubit number, error model, magnitude and optimization level, we simulate $5,000$ QVT$_N$ experiments by sampling $n_c$ circuits from our original sample with replacement. Each sampled circuit has a saved heavy output probability and we perform a binomial sampling with $n_s$ shots to simulate finite sampling effects. We construct the average heavy output frequency of this simulated experiment instance and perform the semi-parametric bootstrap method to calculate the confidence interval lower bound with $n_b=$1,000. Finally, we test to see if this confidence interval lower bound is below the estimated success from the original 5,000 circuit sample to calculate the coverage probability. We repeated this over a grid of experiments with $n_c = [10, 50, 100, 250, 500, 1000]$ and $n_s = [1, 10, 50, 100, 1000]$. We also calculated the original confidence interval for each simulated experiment for comparison.

\begin{figure} 
	\begin{center}
		\includegraphics[width=\columnwidth]{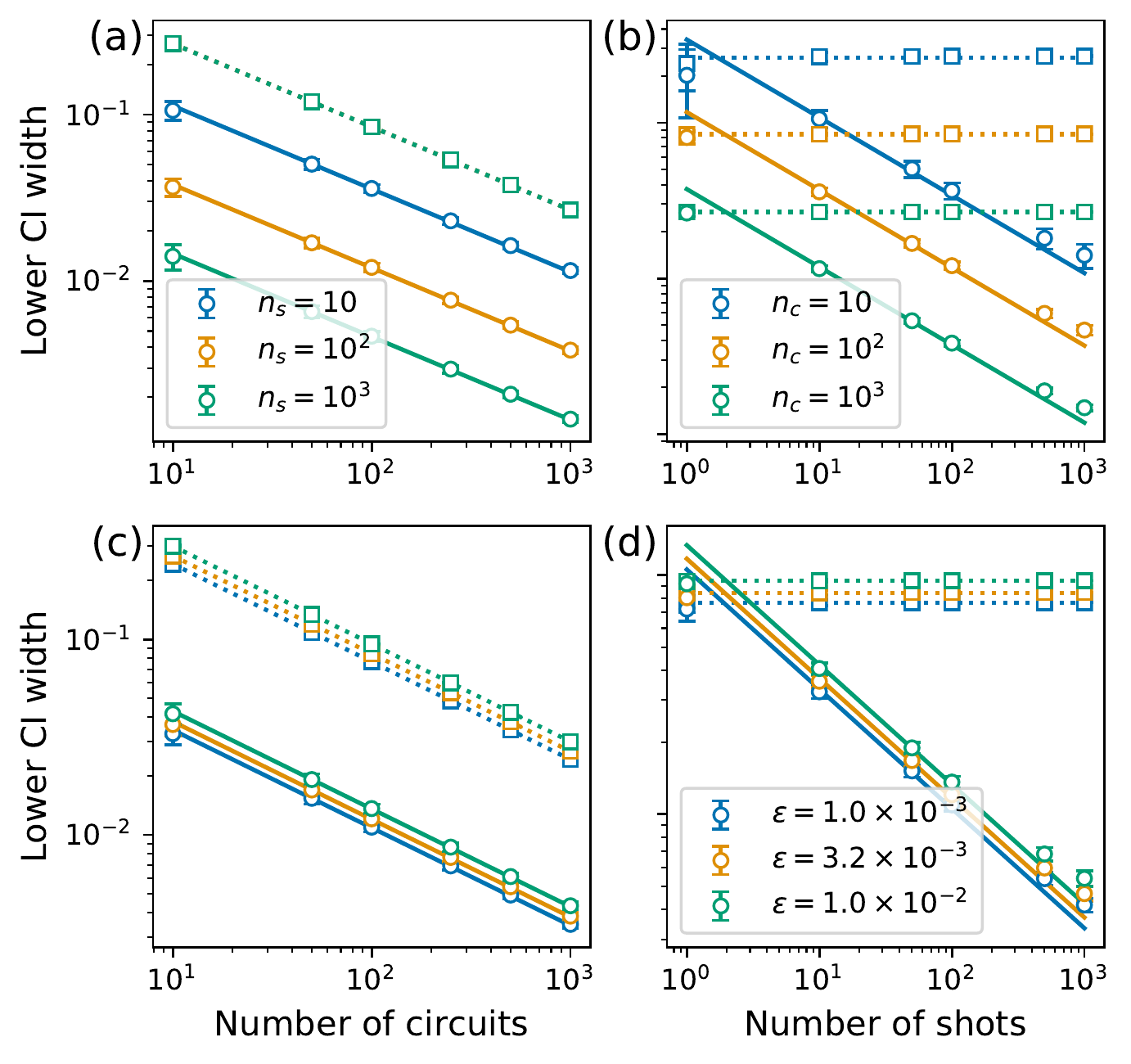}
		\caption{Confidence interval (CI) widths comparison between original method (dashed lines with squares) and bootstrap method (solid lines and circles) for $N=8$ and \textit{Semi-realistic} error model and $high$ optimization. (a) CI width vs. number of circuits for different number of shots. Lines represent fit to $a/\sqrt{n_c}$ with fit parameter $a$. (b) CI width vs. number of shots for different number of circuits. Lines represent fit to $b/\sqrt{n_s}$ with fit parameter $b$. (c) CI width vs. number of circuits for different values of $\varepsilon$ (same legend as d). Lines represent fit to $c/\sqrt{n_c}$ with fit parameter $c$. (d) CI width vs. number of shots for different values of $\varepsilon$. Lines represent fit to $d/\sqrt{n_s}$ with fit parameter $d$. }
		\label{fig:ci_width}
	\end{center}
\end{figure}

The results of the coverage analysis are plotted in Fig.~\ref{fig:distributions2} and show the coverage over different qubit numbers, error models, error magnitudes, sampled circuits and shots all flattened into one dimension. To separate out the effects of shot number we plot the coverage for $n_s=1$ tests separate in Fig.~\ref{fig:distributions2}a and from all other shot numbers in Fig.~\ref{fig:distributions2}b. The dotted horizontal black line shows the specified confidence level 97.73\%. The simulated data shows that both confidence intervals fail to achieve the specified coverage level when $n_s=1$ for most tests (Fig.~\ref{fig:distributions2}a). Using the original method, the lowest coverage occurs for smaller circuit counts ($n_c<100$), which is outside the specifications. However, even for larger $n_c$ the original method still returns coverage around 95$\%$ for several tests. The bootstrap method fails almost uniformly for $n_s=1$.

When going beyond single shot experiments, $n_s>1$, both methods return higher coverage as shown in Fig.~\ref{fig:distributions2}b. The original method has much higher than 97.73\% coverage for all tests and actually achieves unit coverage for most tests. The bootstrap method fails to match the specified coverage for small number of circuits ($n_c = 10$ are plotted as red ``x'') or lower qubit number $N = 4, 5, 6$. However, this should not be a problem when testing $N>6$ and adhering to the QVT requirement of $n_c\geq 100$. We note that the coverage level does seem to increase with qubit number, leaving room for improvement in confidence interval construction for larger $N$.

Larger coverage implies tighter confidence intervals but it is difficult to study how the confidence interval width scales for the bootstrap method since it is numerically estimated and dependent on the error magnitude, $n_s$ and $n_c$. In Fig.~\ref{fig:ci_width} we plot the confidence interval width as a function $n_c$ or $n_s$ with variable $n_c$, $n_s$, or $\varepsilon$ and fixed \textit{Semi-realistic} error model and $N=8$. We see empirically that the width is proportional to $1/\sqrt{n_c}$ in Fig.~\ref{fig:ci_width}a and c but similar attempts to fit the width to $1/\sqrt{n_s}$ do not match the data in Fig.~\ref{fig:ci_width}b and d. The width is also a function of $\varepsilon$ as shown in Fig.~\ref{fig:ci_width}c and d but we did not attempt a fit. Fig.~\ref{fig:ci_width} demonstrates that the bootstrap confidence interval does tighten with number of shots while the original method is constant.

\begin{figure} 
	\begin{center}
		\includegraphics[width=\columnwidth]{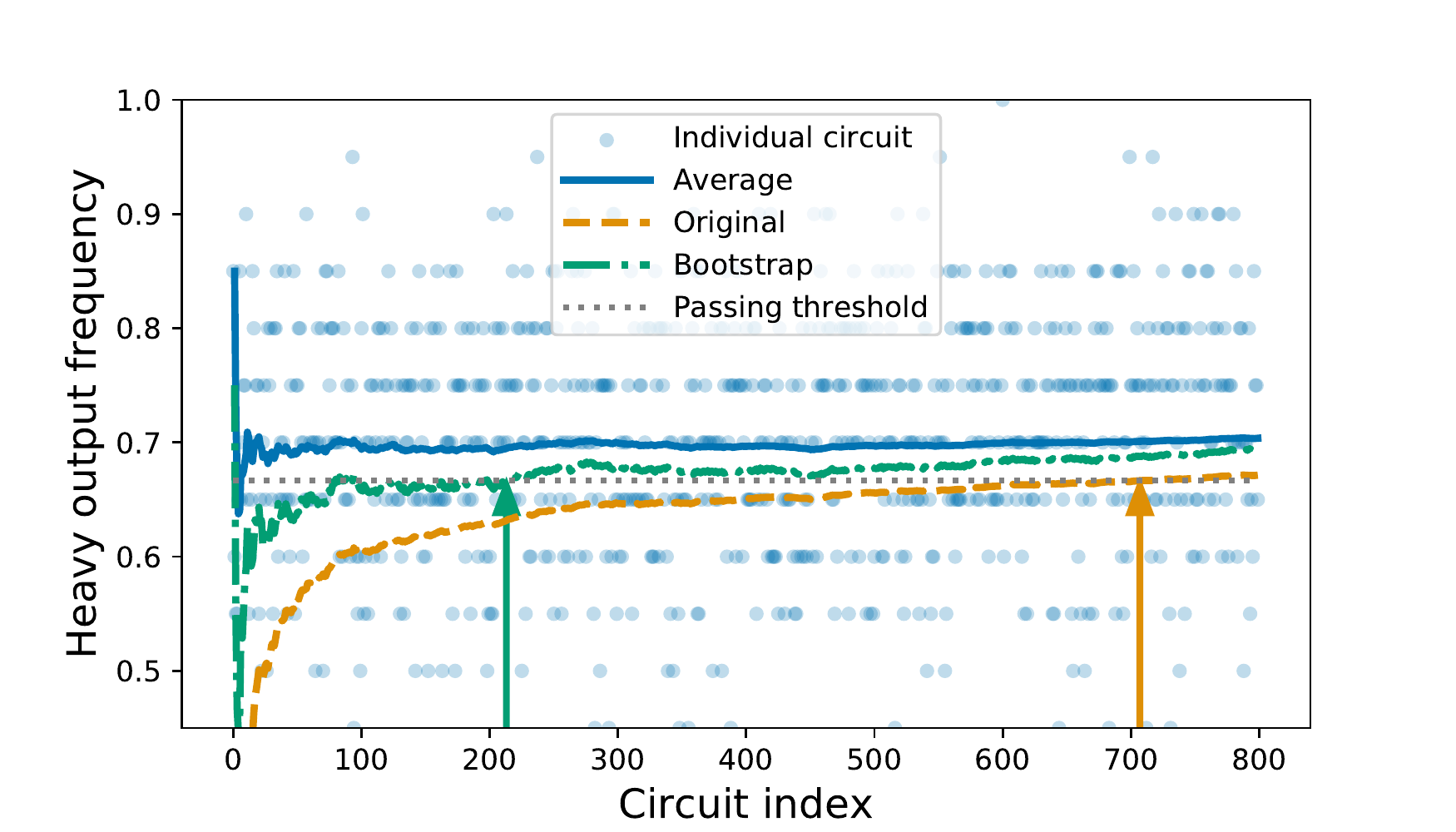}
		\caption{QVT$_{10}$ data from Quantinuum System Model H1-1 with run with $n_c=800$ and $n_s=20$. The final average heavy output frequency is 0.7036. The test is passed with the original confidence interval after 707 circuits (orange arrow) but passes with the new confidence interval after 213 circuits (green arrow) demonstrating the circuit savings of the method.}
		\label{fig:1024_data}
	\end{center}
\end{figure}

As a demonstration of the bootstrapping method we plot the confidence intervals for both methods as a function of number of circuits for the QVT$_{10}$ data announced in Ref.~\cite{HQS1024}. The experiment was performed on the Quantinuum System Model H1-1 machine, similar to the machine discussed in Ref.~\cite{Pino21}. The results are plotted in Fig.~\ref{fig:1024_data} and show that the bootstrap confidence interval method crosses the 2/3 threshold consistently after 213 circuits but the original method crosses at 707 circuits. 

In summary, the original confidence interval construction results in a conservative coverage probability and an excessive circuit number requirement. We constructed a new method to closely match the desired coverage probability, thereby reducing the confidence interval width and saving circuits. We showed that the original design principle of single-shot experiments, $n_s=1$, results in insufficient coverage probabilities for both methods. Our method also converges to higher coverage probability as qubit number increases. Other methods for constructing confidence intervals (or perhaps Bayesian method for credible intervals) might be needed to scale to even larger qubit numbers and handle $n_s=1$ experiments.

\section{Operational implications} \label{sec:comparisons}
The only thing QV perfectly captures, is the ability of a quantum computer to generate the ideal output distributions of random QVT circuits. Relating this ability to other useful tasks necessarily requires assumptions about the noise processes present in the machine under investigation and how those processes impact other algorithms, both of which are typically not well understood. In this section we attempt to relate QVT$_N$ to some near term applications under some assumptions.

\subsection{Random linear depth circuit} \label{sec:circuit_fid}
The QVT$_N$ measured heavy output frequency can also be used as an estimator of average fidelity for $N$-qubit linear depth circuits. This may be useful in relating QVT$_N$ results to state preparation for similar depth quantum circuits. For a QVT$_N$ circuit, we can rewrite the output probability as $p_i = |\bra{x_i} \Lambda U_c \ket{0}|^2$ where $U_c$ is the unitary for a particular QVT$_N$ circuit and $\Lambda$ is the combination of all errors commuted outside of the unitary. In practice, $\Lambda$ is a complicated representation of the errors with $2^{4N}$ parameters (assuming close to best-case Markovian errors~\cite{BlumeKohout21}). Here, we make the oversimplified assumption that $\Lambda$ is a depolarizing error channel, which is similar to our scalable method in Sec.~\ref{sec:scalable}. The exactness of this assumption is an interesting question and it may be reasonable based on the random structure of QVT circuits but we leave that analysis for future work. 

\begin{figure} 
	\begin{center}
		\includegraphics[width=\columnwidth]{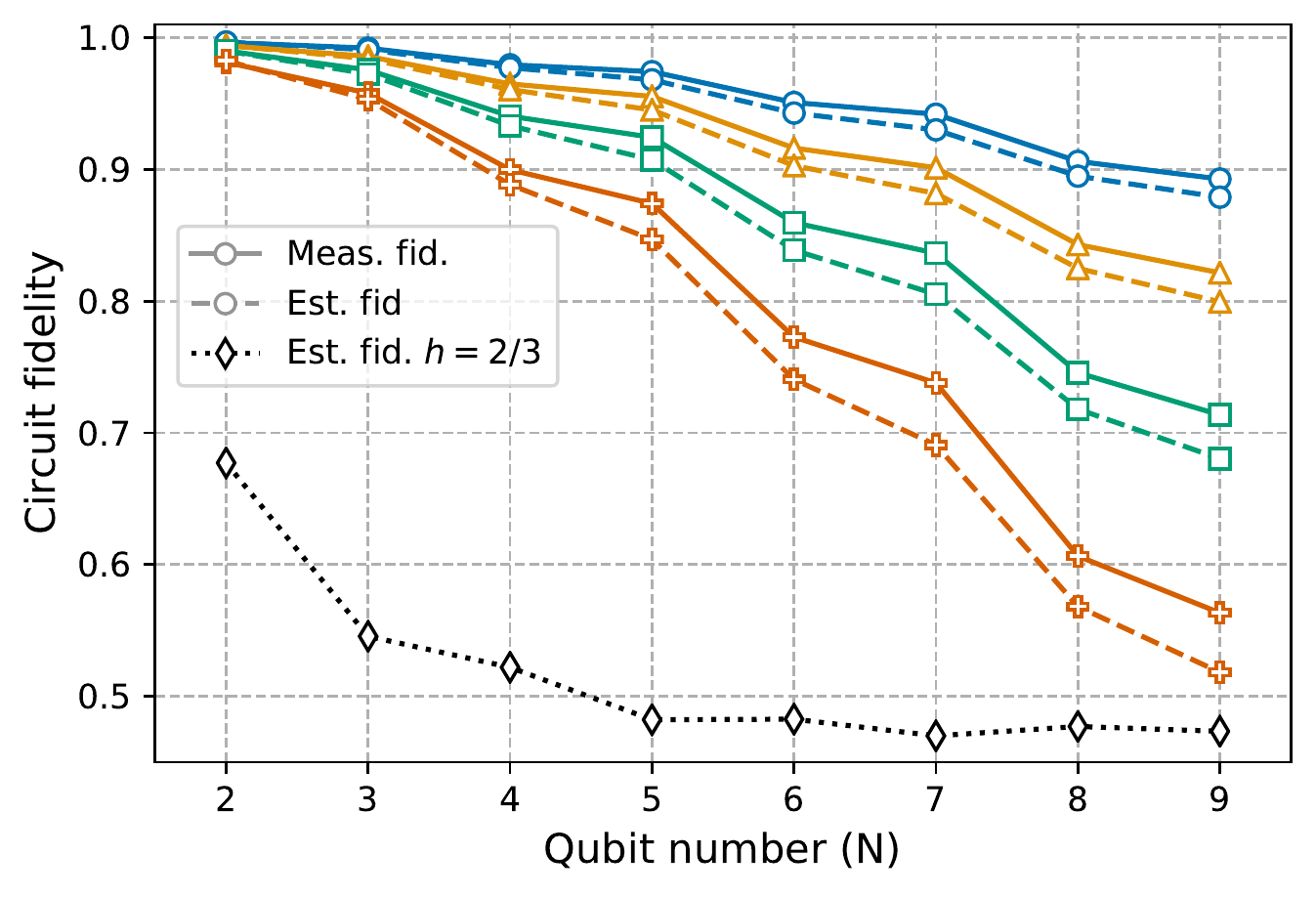}
		\caption{Estimated circuit average fidelity for the \textit{Semi-realistic} model and $high$ optimization with four different error magnitudes: (red crosses) $\varepsilon=1.78\times10^{-2}$, (green squares) $\varepsilon=5.62\times10^{-3}$, (orange triangles) $\varepsilon=1.78\times10^{-3}$, and (blue circles) $\varepsilon=5.62\times10^{-4}$. Solid lines show estimated fidelity from heavy output probability while dashed lines show average state fidelity over the sample. Black dashed line with diamonds show the fidelity corresponding to the passing threshold of 2/3.}
		\label{fig:fidelity_scaling}
	\end{center}
\end{figure}

\begin{table*}[] 
	\centering 
	\resizebox{0.77\paperwidth}{!}{
	\begin{tabular}{cllll}
		\hline
		\hline
		Qubits & Heavy output frequency & Estimated circuit fidelity & Reference     & System                    \\ \hline
		2      & 2/3                   & 0.6771               & -      & -     \\
		2      & 0.718(6)               &0.8086(154)                & Cross \textit{et al.}~\cite{Cross19}      & IBM Tokyo          \\
		2      & 0.7758(417)            & 0.9567(1069)               & Pino \textit{et al.}~\cite{Pino21}       & Honeywell System Model H0 \\ \hline
		3      & 2/3                   & 0.5433              & -      & -     \\
		3      & 0.729(7)               & 0.6997(176)                & Cross \textit{et al.}~\cite{Cross19}      & IBM Johannesburg          \\
		3      & 0.8328(373)            & 0.9603(936)                & Pino \textit{et al.}~\cite{Pino21}       & Honeywell System Model H0 \\ \hline
		4      & 2/3                   & 0.5223               & -      & -     \\
		4      & 0.699(1)               & 0.6114(26)                 & Cross \textit{et al.}~\cite{Cross19}      & IBM Johannesburg          \\
		4      & 0.7677(422)            & 0.8010(1165)               & Pino \textit{et al.}~\cite{Pino21}       & Honeywell System Model H0 \\ \hline
		5      & 2/3                   & 0.4841               & -      & -     \\
		5      & 0.69(1)                & 0.5475(271)                & Sundaresan \textit{et al.}~\cite{Sundaresan20} & IBM Johannesburg          \\ \hline
		6      & 2/3                   & 0.4826             & -      & -     \\
		6      & 0.7296(222)            & 0.6589(622)                & Pino \textit{et al.}~\cite{Pino21}       & Honeywell System Model H0 \\
		6      & 0.701(31)              & 0.579(87)                  & Jurcevic \textit{et al.}~\cite{Jurcevic21}   & IBM Montreal              \\ \hline
		7      & 2/3                   & 0.4708               & -      & -     \\
		7      & 0.7178(159)$^*$            & 0.6129(442)                &   \cite{HQS128}            & Quantinuum System Model H1-1 \\ 
		7      & 0.69(1)$^{*\dagger}$                & 0.54(3)                    &    \cite{IBM128}           & IBM Montreal              \\ \hline
		9      & 2/3                   	   & 0.4737               & -      & -     \\
		9      & 0.7332(255)$^*$           & 0.6620(723)              &     \cite{HQS512}          & Quantinuum System Model H1-1 \\ 
		\hline
		10      & 2/3                  	   & 0.4788               & -      & -     \\
		10     & 0.7036(102)          & 0.5846(293)              &     Fig.~\ref{fig:1024_data}, \cite{HQS1024}         & Quantinuum System Model H1-1 \\
		\hline
		\hline
	\end{tabular}
	}
	\caption{Heavy output frequency and estimated circuit fidelity of all experimentally passed QVT$_N$ with reported values as of writing. Uncertainty is reported based on the original confidence interval in Ref.~\cite{Cross19} since it is the only estimate available for most data; however, it is generally larger than what we find from our new method in Sec.~\ref{sec:confidence}. For each dataset we use the estimated value of $h_{\textrm{ideal}}(N)$ from numerical simulations in Sec.~\ref{sec:ideal}. $^*$Data was provided by company and not published or in preprint at time of writing. $^{\dagger}$Value is estimated from referenced plot but not confirmed.} \label{table2}
\end{table*}

For a full-circuit depolarizing channel the heavy output probability of a given circuit is directly related to the circuit's depolarizing parameter. From the depolarizing parameter, we calculate the average circuit fidelity,
\begin{equation}
F_{\textrm{circ}} = 1 - \frac{2^N - 1}{2^N}\frac{h_{\textrm{ideal}}(N) -\hat{h}}{h_{\textrm{ideal}}(N) - 1/2},
\end{equation}
where $h_{\textrm{ideal}}(N)$ is the heavy output probability without errors (studied in Sec.~\ref{sec:ideal}) and $\hat{h}$ is the heavy output frequency with errors for a given circuit. This is similar to the quantity proposed in Refs.~\cite{Kim21, Liu21} but with a dimensional scaling factor. A totally depolarized circuit has $F_{\textrm{circ}} = 1/2^N$. The QVT passing threshold of 2/3 corresponds to $F_{\textrm{circ}} =1/3 \ln 2 \approx0.481$ in the asymptotic limit of large $N$ but is in general a function of $N$. 

In Fig.~\ref{fig:fidelity_scaling}, we compare the estimated circuit fidelity $F_{\textrm{circ}}$ to the average state fidelity of the output averaged over 5,000 simulated QVT$_N$ circuits, which we use as an approximation of the average fidelity of the QVT$_N$ circuits. We study the \textit{Semi-realistic} model and $high$ optimization with four different error magnitudes. The estimated fidelity from the heavy output probability consistently overestimates the average fidelity. This is contrary to a similar studies performed in Ref.~\cite{Liu21}, which uses different circuit construction that produce estimates that closely matches the fidelity. Further investigation is required to understand why QVT circuits slightly overestimate fidelity.

As shown in Sec.~\ref{sec:ideal}, the ideal output states of the QVT$_N$ for large $N$ are highly entangled. Therefore, we can use the estimate $F_{\textrm{circ}}$ as an estimate for the fidelity of entangled state preparation with comparable depth circuits. Entangled state preparations are important in several near term algorithms such as VQE~\cite{McClean16}. For reference, Table~\ref{table2} shows the conversion of recent QVT$_N$ data to fidelity estimates. For this table we used the average heavy output frequency over all circuits and the expected ideal heavy output probability from Sec.~\ref{sec:ideal} instead of a per circuit estimate.

\subsection{Quantum error correction} \label{sec:qec}
It is widely believed that quantum computers will require quantum error correction (QEC) to reach error rates necessary to perform large-scale quantum computation~\cite{devitt2013quantum}. QEC works by encoding quantum information into logical qubits, which are constructed from many physical qubits, with a QEC code. There are several different proposals for QEC codes but they all use physical qubits to detect and correct certain errors in the logical qubits without destroying the underlying quantum information. A QEC code is partially defined by its distance $d$, which roughly indicates the number of physical errors a code can tolerate and correct. Broadly speaking, the number of qubits needed to implement a QEC code grows polynomially with the code's distance. QEC consists of structured and repetitive circuits, seemingly quite different than QVT circuits. Here, we compare a QEC code's logical pseudo-threshold to QVT$_N$ passing thresholds. 

We define the pseudo-threshold of a QEC code as the point where the logical error rate is equal to the largest physical error rate. For example, with the \textit{TQ Depolarizing} model, the pseudo-threshold is the point that the logical error rate is equal to the two-qubit depolarizing error rate. Ideally, a system implementing a QEC code will operate well below the pseudo-threshold to take advantage of the error suppression. In fact, in a real system the pseudo-threshold will be difficult to measure since it requires full knowledge of all error sources. However, in simulations it can be well defined, and it is a natural performance metric of different codes to compare with QVT requirements.

To probe the relationship between a system's ability to pass QVT$_N$ and its expected performance of QEC codes, we ran simulations for three different QEC codes: the [[7,1,3]] Steane code~\cite{steane1996simple}, the rotated surface code~\cite{fowler2012surface}, and the toric code~\cite{bravyi1998quantum}. We applied three error models introduced in Sec.~\ref{sec:error_channels}: \textit{TQ depolarizing}, \textit{Measurement}, and \textit{Memory}. For the [[7,1,3]] Steane code, the simulation was done using a stabilizer simulator with a lookup table style decoder~\cite{ryan2018performance} and for the rotated surface and toric codes, a fast Pauli tracking simulator was used with minimum weight perfect matching to decode error syndromes~\cite{edmonds1965paths, fowler2014minimum}. These choices allow us to investigate the relationship between QVT and QEC for both low and higher distance codes. 

\begin{figure} 
	\begin{center}
		\includegraphics[width=\columnwidth]{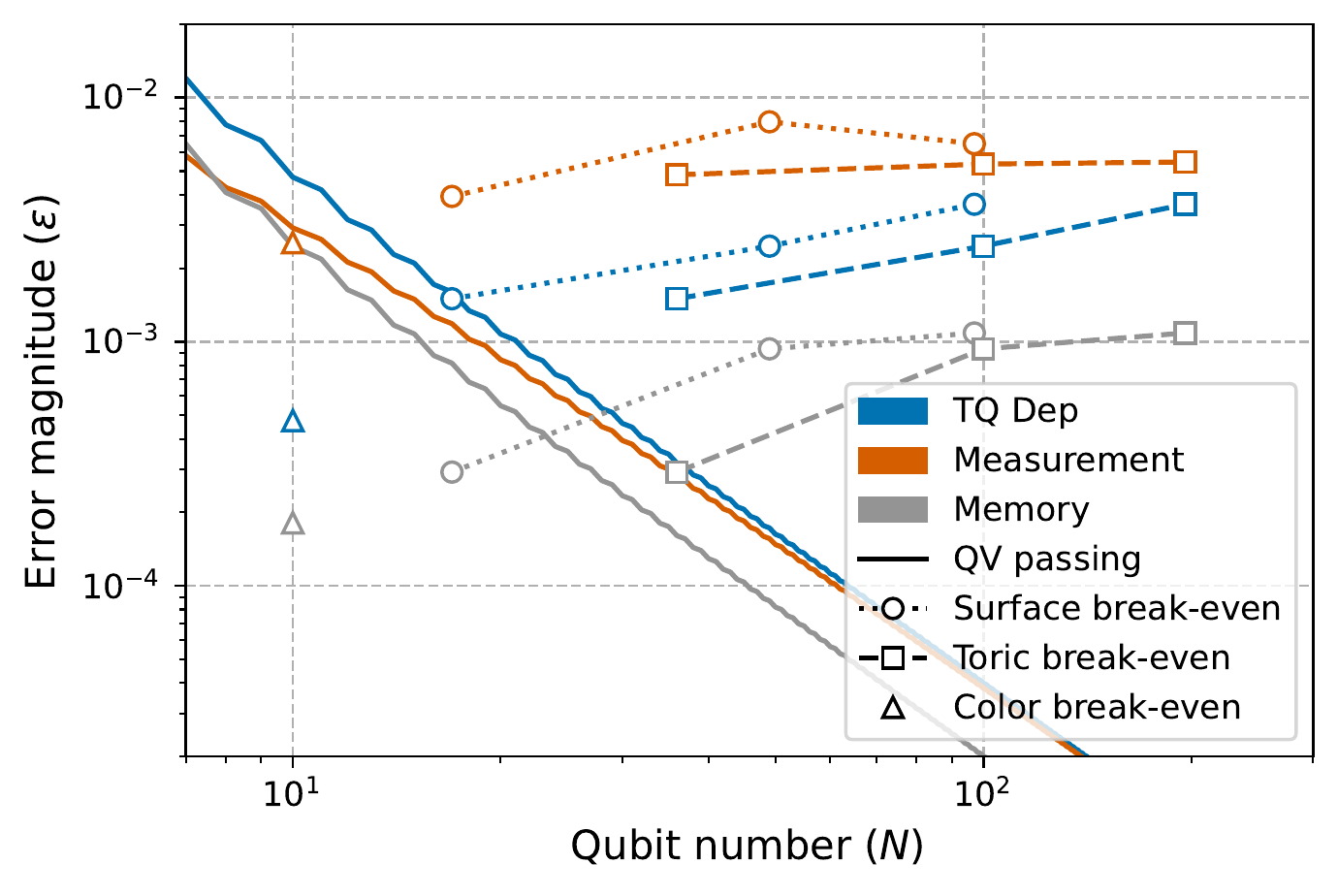}
		\caption{QVT passing error magnitudes compared to QEC pseudo-thresholds with three different error models. Colors correspond to error models and line-styles and labels correspond to different thresholds and codes. The QVT passing threshold is estimated from the scalable model with the \textit{dep} option (solid lines). The three QEC codes tested are: rotated surface code for $d=3,\:5,\text{\;and\:}7$ (circles with finely dashed lines), toric code for $d=3,\:5,\text{\:and\:}7$ (squares with dashed lines), and  [[7,1,3]] Steane code (triangles). Three error models were tested are \textit{TQ depolarizing} (blue), \textit{Measurement} (red), and \textit{Memory} (grey). As qubit number increases, the error magnitude required to pass QV exponentially decays, whereas the error magnitude required for a QEC code to reach the pseudo-threshold increases.}
		\label{fig:QEC}
	\end{center}
\end{figure}

Fig.~\ref{fig:QEC} shows the error magnitude required to pass QVT$_N$ compared to the error magnitude required to reach the pseudo-threshold for the Steane code with $d=3$, the rotated surface code with $d=3,\:5,\text{\:and\;}7$, and the toric code with $d=3,\:5,\text{\:and\:}7$.  The estimated QVT$_N$ passing thresholds does match the pseudo-threshold for a few codes and error models (e.g., distance three surface with the \textit{TQ depolarizing} model and Steane with the \textit{Measurement} model). However, most psuedo-threshold points fall outside of the QVT$_N$ passing estimates. Therefore, we do not find that passing QVT$_N$ implies being able to reach the pseudo-threshold for a specific QEC code. 

QVT$_N$ may not be predictive for a specific code but the qubit and fidelity requirements do roughly align well with a general class of small-distance codes. The pseudo-thresholds for distance three codes roughly fall within a region of $9< N <30$ and $4\times10^{-2} \leq \varepsilon  \leq 2\times10^{-4}$ for all error models tested and the QVT$_N$ passing thresholds intersect this region. Designing and verifying that a machine can pass QVT$_N$ within this region would provide a reasonable starting point for testing a variety of small-distance codes. 

Furthermore, since QVT$_N$ requires arbitrary connectivity then passing QVT$_N$ within the region defined above implies many codes are available to test. We simulated example codes with nearest-neighbor parity measurements but in principle one may want to test other codes, such as LDPC codes using non-local parity checks~\cite{Babar15} or randomly generated codes~\cite{Gullans21}. Verifying low error rate connections means such codes should be feasible to implement and compare.

Scaling QVT$_N$ to larger $N$ necessarily requires lowering error rates but scaling QEC to larger distances, which also requires more qubits, actually alleviates requirements on error rates to reach pseudo-thresholds. This is shown with the rotated surface and toric codes in Fig.~\ref{fig:QEC}, which have pseudo-thresholds error magnitudes which increase (are easier to achieve) with larger $N$. Thus there is a crossover regime after which passing QVT$_N$ becomes more difficult than reaching the pseudo-threshold for a QEC code with the same number of qubits. Lowering error rates is always beneficial for QEC but not strictly necessary after the crossover regime. 

QEC requires additional features that are not necessary to run QVT. In order to implement QEC a quantum computer needs to be able to apply mid-circuit measurements and resets to measure errors and, ideally, feed-forward operations to correct errors~\cite{RyanAnderson21}. QVT does not require either of these features so even if a quantum computer can implement and pass QVT$_{30}$, for example, it will not necessarily be able to run QEC. 

Another notable difference between QVT and QEC is the dependence on type of error. While we have not simulated every type of error for QVT we did observe in Sec.~\ref{sec:numerics} that many different types of errors have similar effects on the passing threshold of QVT. However, the same is not true for QEC where some types of errors cannot be corrected in certain codes. For example, leakage errors map population outside the qubit manifold and can rapidly accumulate and spread additional errors, breaking the fault tolerance of the code \cite{suchara2015leakage, brown2018comparing}. Coherent errors are known to have a larger impact on smaller distance codes compared to larger distance codes~\cite{iverson2019coherence}. Proving an implementation of a QEC code is below the pseudo-threshold requires probing several different basis states to confirm an arbitrary state is also below the pseudo-threshold~\cite{gottesman2016quantum}. Any asymmetry in the errors will affect the basis states differently and possibly cause certain states to not meet pseudo-threshold.

We did not compare such errors because in practice they can be mitigated by circuit design~\cite{brown2020critical, brown2019handling, zhang2021hidden, ParradoRodrguez2021, Tuckett18, Wallman16, debroy2018stabilizer} or physical techniques~\cite{hayes2020eliminating}. Moreover, we do not anticipate these errors to be the leading order errors in near-term systems where QVT is applicable. We leave such detailed comparisons to future work.

Presently, both QVT and QEC demonstrations are well aligned with near term goals of increasing qubit number and decreasing error rates. However, as devices continue to mature QVT tests will no longer be classically simulatable and also harder to pass while QEC will necessarily be required to scale to larger qubit numbers.

\section{Conclusions} \label{sec:conclusions}
Our work illuminates previously unstudied behavior of QVT and requirements for scaling to larger $N$. We first considered how circuit construction impacts the test results. Even without errors the ideal heavy output probabilities scale with the qubit number, which has a notable impact for $N<10$. The standard optimizations used on the circuits also have a significant effect for $N<10$, which reduces two qubit gates by at least 20\%, but is less effective as $N$ increases. Next, we preformed a series of simulations to test the behavior of QVT$_N$ with different error sources. We observed that QVT$_N$ passing is mostly dependent on total error magnitude (including errors like crosstalk) and not on the error sources.  We constructed a scalable method for larger qubit numbers that roughly estimates error requirements for passing QVT$_N$. After, we studied the confidence interval construction for QVT$_N$ from Ref.~\cite{Cross19} and found that the method returned much higher coverage than specified for most experiments except when run with a single shot per circuit where the coverage was lower than specified. We proposed a new method that returns tighter confidence intervals and showed it had near the expected coverage with more than one shot and 100 circuits. Finally, we compared QVT$_N$ results to other important quantum computing applications. We showed that the heavy output probability can be converted to serve as an estimate that scales with the average state preparation fidelity although is generally slightly higher. We also numerically demonstrated that the fidelity requirements for QVT$_N$ for $9 < N  < 30$ roughly align with the fidelity requirements for many low-distance break-even QEC demonstrations. There is one obvious question left out of the FAQ's for QVT in Sec.~\ref{sec:faq}; ``Is QVT a good benchmark?'' This is clearly a complicated question with a variety of opinions but we observe that most disagreements come down to two main questions about full system benchmarking:  

\Q{Is random circuit construction a reasonable way to benchmark systems?} 
\A{The random circuit construction for QVT captures the effects of many different error sources (as seen in Sec.~\ref{sec:numerics}) but lead to previously unknown irregular performance effects dependent on qubit number --- as shown in Sec.~\ref{sec:circuits}. Capturing different error sources is crucial to near-term benchmarking since many systems suffer from errors that are missed in individual component benchmarks and usually not well understood. The irregular performance diminishes for larger qubit number and do not seem likely to be a problem for future QVT$_N$ experiments ($N>10$). Another downside we identified is that QVT seems to mostly scale with the total error magnitude (infidelity). This may mean some errors, like coherent errors, may have different effects in QVT than in certain algorithms. This is a typical downside of random circuit benchmarking but allows the results to better relate to circuit fidelity (as studied in Sec.~\ref{sec:circuit_fid}). Finally, QVT requires high-fidelity interaction between arbitrary qubits, which is not required for many algorithms and QEC codes. Nevertheless, such high-fidelity arbitrary interactions are useful for near-term devices to test and compare a variety of algorithms or QEC codes, which may have widely different connectivity requirements.}

\Q{Are square circuits the best choice for judging a quantum computer's performance?} 
\A{While QVT circuits have linear depth, the fidelity requirements to pass QVT$_N$ for $9<N<30$ match well with other goals for quantum computation, especially QEC. Non-Markovian system errors that occur in longer circuits are missed in QVT$_N$, which is one downside to the test, but square scaling balances fidelity and qubit requirements in a reasonable way for near-term goals.}

Overall, we believe our work supports the notion that QVT is a good benchmark, with the above caveats, but QVT is certainly not the only or final answer to full system benchmarking of quantum computers. In practice, it is best to use multiple benchmarks that stress different circuit sizes and errors to fully judge a systems performance. Ultimately, we expect a suite of benchmarks with comprehensive studies --- like we attempted here --- will serve as standards for comparing different systems. Moreover, QVT in its current form will not be useful for more than $\sim 30$ qubits and as platforms move towards QEC the need to scale qubit number will outweigh the need to scale fidelity. However, we find that currently QVT does set worthwhile near-term goals for performance demonstrations that measure system level errors.

\begin{acknowledgments}
We would like to thank the entire Quantinuum team for helpful input and questions that inspired this work. We especially thank the Quantinuum System Model H0 and H1 teams and other contributors for running the quantum volume tests. Specifically Dan Gresh, Aaron Hankin, Kevin Gilmore, Justin Gerber, John Gaebler, David Francois, Thomas Gatterman, Si Khadir Halit, Alex Hall, Justin Bohnet and Brian Neyenhuis who contributed to the QVT$_{10}$ data presented.
\end{acknowledgments}

\end{document}